\documentclass[final,3p,times]{elsarticle}

\usepackage{graphicx}

\usepackage{amssymb}

\usepackage[english,francais]{babel}

\def\og{\leavevmode\raise.3ex\hbox{$\scriptscriptstyle\langle\!\langle$~}}
\def\fg{\leavevmode\raise.3ex\hbox{~$\!\scriptscriptstyle\,\rangle\!\rangle$}}

\usepackage[none]{hyphenat}
\usepackage[T1]{fontenc}
\usepackage{amsmath}
\usepackage{color}
\newcommand{\bea}{\begin{eqnarray}}
\newcommand{\eea}{\end{eqnarray}}
\newcommand{\be}{\begin{equation}}
\newcommand{\ee}{\end{equation}}
\newcommand{\rr}{\mathbf{r}}
\newcommand{\kk}{\mathbf{k}}

\newcommand{\PP}{\mathbf{P}}
\newcommand{\pp}{\mathbf{p}}
\newcommand{\qq}{\mathbf{q}}

\newcommand{\vv}{\mathbf{v}}
\newcommand{\kf}{k_{\mathrm{F}}}

\usepackage{float}

\begin{document}

\begin{frontmatter}


\selectlanguage{francais}
\title{The Landau critical velocity for a particle in a Fermi superfluid}


\author[lkb]{Yvan Castin}
\author[lkb]{Igor Ferrier-Barbut}
\author[lkb]{Christophe Salomon}

\address[lkb]{Laboratoire Kastler Brossel, ENS-PSL, CNRS, UPMC-Sorbonne Universit\'es and Coll\`ege de France, Paris, France}



\begin{abstract}
We determine \`a la Landau the critical velocity $v_c^{L}$ of a moving impurity 
in a Fermi superfluid, that is by restricting to the minimal excitation processes 
of the superfluid.
$v_c^{L}$ is then the minimal velocity at which these processes are energetically allowed. The Fermi superfluid actually exhibits two excitation branches~: one is the fermionic
pair-breaking excitation, as predicted by BCS theory; the other one is bosonic and sets
pairs into motion, as predicted by Anderson's RPA. $v_c^{L}$ is the smallest
of the two corresponding critical velocities $v_{c,f}^{L}$ and $v_{c,b}^{L}$.
In the parameter space (superfluid interaction strength, fermion-to-impurity mass ratio),
we identify two transition lines, corresponding to a discontinuity of the first-order
and second-order derivatives of $v_c^{L}$. These two lines meet in a triple point and split
the plane in three domains. We briefly extend this analysis to the very recently realized
case at ENS, where the moving object in the Fermi superfluid is a weakly interacting
Bose superfluid of impurities, rather than a single impurity. 
For a Bose chemical potential much smaller than the Fermi energy, the topology of
the transition lines is unaffected; a key result is that the domain $v_c^{L}=c$,
where $c$ is the sound velocity in the Fermi superfluid, is turned into a domain
$v_c^{L}=c+c_B$, where $c_B$ is the sound velocity in the Bose superfluid, 
with slightly shifted boundaries.
\noindent 
\vskip 0.5\baselineskip

\selectlanguage{english}
\end{abstract} 
\end{frontmatter}

\selectlanguage{english}

\section{Introduction, reminders et motivations}
\label{sec:intro}

Degenerate gases of interacting spin $1/2$ fermionic neutral atoms, taken here to be non-polarized that is with equal populations
in the two internal states, have been realized in the laboratory since 2002 \cite{manips_fermions}.
Below a critical temperature, they exhibit two distinct and remarkable macroscopic quantum properties.
The first one is the presence of a condensate of pairs, that is the existence of a macroscopically populated mode
of the two-body density operator \cite{Leggett_POF}, which physically implies a macroscopic coherence length 
for the field of pairs, only limited by the size of the system. This ``long-range order" can be in principle be directly
measured by interferometry \cite{Iacopo_interf}, but only the fraction of condensed pairs $f_c$ was measured up to now \cite{fcpaires}.
The second property, the one that is of interest here, is superfluidity. It is perceived as being more subtle, because it involves
a complex of phenomena, some of them relying on metastability rather that on equilibrium properties. Let us mention here
only two aspects, skipping quantum vortex lattices \cite{Zwierlein_tourb} and permanent currents.

The first aspect involves the concept of superfluid fraction $f_s$: for periodic boundary conditions in a cube of size $L$,
it is the fraction of the gas which is not set into motion by a moving external potential, even after an arbitrarily long time
allowing the system to reach thermal equilibrium in the moving frame. For an external potential moving along direction $x$
at velocity $v$, the normal fraction $f_n=1-f_s$ of the gas is by definition moving at that velocity, so that
\be
1-f_s= \lim_{N\to\infty,\,\rho=\mathrm{ const}} \ \ \lim_{v\to 0} \ \ \lim_{\eta\to 0} \ \ \frac{\langle P_x\rangle}{N m v}
\label{eq:def}
\ee
where $N$ is the number of atoms in the gas, $m$ is the atom mass, $\rho=N/L^3$ is the gas density and
$\langle P_x\rangle$ is the mean equilibrium total momentum of the gas along $x$ in presence of the external potential.
The triple limit must be taken in that order, so as to make the normal fraction an intrinsic quantity.
On first takes the limit of a vanishing external potential amplitude $\eta$, so that $f_n$ does not depend on the shape of the potential.
Then on takes the limit of a vanishing stirring velocity, {\sl before} one takes the thermodynamic limit, so as to always have
\be
v \ll \frac{2\pi \hbar}{mL}
\label{eq:condv}
\ee
For $v$ equal to the velocity quantum $2\pi\hbar/(mL)$, the periodic boundary conditions are indeed Galilean invariant, so that
the gas would be at rest in the external potential frame and would have a mean momentum $\langle P_x\rangle=N m v$
in the lab frame, leading to the constant (and unphysical) result $f_n=1$ { \cite{Caruso}}.
The superfluid fraction $f_s$ of a spin $1/2$ non-polarised Fermi gas was very recently measured in the strongly interacting regime,
as a function of temperature $T$ \cite{Grimm}, allowing one to check that the superfluid phase transition takes place
at the pair condensation temperature \cite{fcpaires} and at the temperature where thermodynamic quantities have singularities \cite{ENSZwierlein}.
An expected key property is that $f_s\to 1$ at zero temperature, and this is experimentally confirmed on other systems.

The second aspect of superfluidity is in principle restricted to the zero temperature case. It corresponds to the existence of a critical velocity
$v_c$ below which an object moving through the gas does not experience any friction force and cannot deposit any energy, thus having an
undamped motion. This aspect was indeed observed with cold atomic Fermi gases for a moving one-dimensional optical lattice \cite{super_spectacle}.
Predicting the critical velocity is often difficult, as it generally depends on the object intrinsic properties and on its coupling to the gas
\cite{Pomeau}. For an arbitrarily weak coupling to the gas density however\footnote{This can be an effective coupling: for a pointlike object, it is
proportional to its $s$-wave scattering amplitude with the atoms of the gas.},
in the spirit of the definition (\ref{eq:def}), one may limit oneself, as Landau did for a Bose gas \cite{Landau}, to the first step 
in the dissipation of the object kinetic energy, that is the creation of the minimal number of elementary excitations in the gas,
a single excitation in the case of \cite{Landau}. Formally, this amounts to evaluating the object-gas scattering amplitude in the Born
approximation, to first order in the gas-object coupling, or the emission rate of excitations by the object according to the Fermi golden
rule, to second order in the coupling constant; in both cases, one gets as a factor a Dirac distribution ensuring the conservation of
unperturbed energy \cite{Pitaevskii}. In this work, except in section \ref{sec:cbe}, the object is a particle of mass $M$, distinguishable
from the atoms of the gas, of initial velocity $v$ and initial kinetic energy $\frac{1}{2}M v^2$. After emission in the gas of an excitation
of wavevector $\qq$ and energy $\epsilon_\qq$, its velocity is $\mathbf{v} -\hbar\qq/M$, due to momentum conservation, so that
\be
\hbar \qq\cdot \vv = \frac{\hbar^2 q^2}{2M} +\epsilon_\qq
\ee
due to unperturbed energy conservation. Since $|\qq\cdot \vv|\leq qv$, this condition cannot be satisfied by any
$\qq$ if $v$ is below the Landau critical velocity
\be
v_c^{L} = \inf_\qq v_\qq \ \ \ \mbox{with} \ \ \ v_\qq=\frac{\frac{\hbar^2q^2}{2M} + \epsilon_\qq}{\hbar q}
\label{eq:vcL}
\ee
For a particle of mass $M\to+\infty$, $v_c^{L}$ was calculated in \cite{CKS} with the approximate BCS and RPA theories;
to this end, both excitation branches of the Fermi superfluid were taken into account, the gapped fermionic branch
corresponding to a breaking of Cooper pairs of atoms, and the gapless bosonic branch with a  phononic behavior close to
$\qq=\mathbf{0}$, corresponding to pairs being set into motion. One then gets \cite{CKS}
\be
v_c^{L}(\alpha=0) = \min(\{[(\mu^2+\Delta^2)^{1/2}-\mu]/m\}^{1/2},c)
\label{eq:cks}
\ee
where $c$ is the sound velocity in the Fermi superfluid, $\mu$ is the chemical potential, $\Delta$ is the gap, and
the fermion-impurity mass ratio is denoted by
\be
\alpha = \frac{m}{M}
\label{eq:defalpha}
\ee
The predicted critical velocity for $M\to +\infty$ is different from zero. This seems to contradict the reasoning below 
equation (\ref{eq:condv}). The effect of an infinite mass object on the superfluid is the same as an external potential 
moving at constant velocity, so that one should have $v_c \leq  2\pi \hbar/(mL) \to 0$ in the thermodynamic limit.
Landau's reasoning is however saved by the (subtle) concept of metastability, which gives a physical meaning to the
predicted critical velocity $v_c^{L}$ at least at short times: for $v< v_c^{L}$, the first step towards dissipation
is blocked by an energy barrier, but the system can in principle overcome this barrier at long times thanks to processes of arbitrarily high
order in the gas-object coupling, involving an excitation energy $\epsilon_\qq$ very different from the one of the elementary excitations
of equation (\ref{eq:vcL}). For example, the improbable process of order $N$ setting the whole gas into motion at velocity
$2\pi \hbar/(mL)$ along $x$, by a momentum boost of $2\pi\hbar/L$ applied to each atom along that direction, corresponds
to $q=2\pi N/L$ and $\epsilon_\qq=N (2\pi\hbar)^2/(2mL^2)$, that is to a critical velocity $v_\qq=2\pi\hbar/(2 m L)$
that indeed vanishes in the thermodynamic limit
\footnote{One more commonly invokes the vortex ring as the macroscopic excitation created by the moving object in the gas. As a function
of its radius $R$, it has an energy scaling as $R\ln R$ and a momentum scaling as $R^2$ \cite{Svistunov_livre}.
When $R$ reaches the gas diameter, one is led to the same $N$ and $L$-scaling laws for $q$, $\epsilon_\qq$ et $v_\qq$,
up to a factor $\ln L$ in $\epsilon_\qq$. If the object is a pointlike particle weakly coupled to the superfluid, the emission of a vortex ring 
remains anyway improbable at velocities below $v_c^L$ \cite{Pitaevskii}.}
provided that $M$ remains $\gg N m$. \footnote{For a finite mass $M$ and a sufficiently weak repulsive coupling between the impurity and the superfluid, it was
recently shown that there exists a genuine critical velocity $v_c$ that does not vanish in the thermodynamic limit, that does not
rely on metastability considerations and holds for an arbitrarily long interaction time between the impurity and
the superfluid \cite{Lychkovskiy}. When the coupling tends to zero, $v_c$ is obtained by including in (\ref{eq:vcL}) all
the possible superfluid excitations, and not only the elementary excitations as it will be done here. For the vortex
ring of the previous footnote and after minimisation of $v_\qq$ over the ring radius, one finds that $v_c$ vanishes
as $(\ln M)^{2/3}/M^{1/3}$ when $M\to +\infty$.}

The goal of the present work is to extend the calculations of reference \cite{CKS} to the case of a finite mass $M$.
There is a very strong experimental motivation: the Landau prediction for the critical velocity of an atomic impurity
in a Bose-Einstein condensate was confirmed at MIT \cite{Ketterle_Landau}, and may be measured soon in a Fermi superfluid
thanks to the Bose-Fermi mixture of superfluids recently obtained at ENS \cite{Salomon_melange}.
The calculation of $v_c^{L}$ for a particle of mass $M$ is performed here in three steps:
one determines the critical velocity $v_{c,f}^{L}$ due to the BCS fermionic excitation branch in section \ref{subsec:vcf},
then the one  $v_{c,b}^{L}$ due to the RPA bosonic branch in section \ref{subsec:vcb}, then one takes the smallest of the two velocities
to get $v_c^{L}$ in section \ref{subsec:vc}.
In reference \cite{Salomon_melange} the object moving inside the Fermi superfluid is a Bose-Einstein condensate rather than 
a single impurity, so that we modify Landau's formula (\ref{eq:vcL}) in section \ref{sec:cbe} to take into
account the interaction among impurities. We conclude in section \ref{sec:conclusion}.

\section{Critical velocity on the fermionic branch}
\label{subsec:vcf}

One might naively believe that the critical velocity $v_{c,f}^{L}$ associated with the fermionic superfluid excitation
branch is derived from the general expression (\ref{eq:vcL}) by taking for $\kk\mapsto \epsilon_\kk$ the corresponding dispersion relation
$\kk\mapsto \epsilon_{f,\kk}$ of the fermionic quasiparticles.
However, this would be wrong because it would ignore the constraints imposed by
the conservation of the total number of fermions. In reality, the impurity of mass $M$,
interacting with the superfluid initially in the vacuum of quasiparticles, can produce only an {\sl even} number
of fermionic excitations.
This is particularly clear in the context of BCS theory: in second quantized form, the two-body interaction Hamiltonian 
between the impurity and the fermions involves the fermionic fields $\hat{\psi}_\sigma(\rr)$, $\sigma=\pm 1/2$ 
only through quadratic terms of the form $\hat{\psi}_\sigma^\dagger \hat{\psi}_{\sigma'}$;
but each $\hat{\psi}_\sigma(\rr)$ is a linear combination of quasi-particle annihilation $\hat{b}_{\kk\sigma}$ and 
creation $\hat{b}_{\kk-\sigma}^\dagger$ operators that change the parity of their number.
In Landau's reasoning, it must then be assumed that the impurity creates at least {\sl two} fermionic quasiparticles,
of wavevectors $\kk_1$ and $\kk_2$. The impurity then experiences a momentum change of $-\hbar(\kk_1+\kk_2)$ 
and acquires a recoil energy of $\hbar^2(\kk_1+\kk_2)^2/(2M)$, which leads to the critical velocity
\footnote{In the case of a moving object of infinite mass, one finds however in the literature the usual 
formula $v_{c,f}^{\mathrm{hab}}=\inf_q \epsilon_{f}(q)/(\hbar q)$ \cite{CKS}, with $\epsilon_f(q)=\epsilon_{f,\qq}$,
which seems to correspond to the naive error mentioned above. In reality, our equation (\ref{eq:vcfun}) correctly
reproduces $v_{c,f}^{\mathrm{hab}}$ when $M\to+\infty$. First, ${\lim_{M\to +\infty}} v_{c,f}^{L}\leq v_{c,f}^{\mathrm{hab}}$
since equation (\ref{eq:vcfun}) contains $\kk_1=\kk_2$ as particular configurations. Second, minimisation over the directions of
$\kk_1$ and $\kk_2$ for fixed moduli is straightforwardly achieved when $M=+\infty$ for parallel wavevectors, so that ${\lim_{M\to +\infty}} v_{c,f}^{L}
=\inf_{k_1,k_2}\frac{\epsilon_f(k_1)+
\epsilon_f(k_2)}{{\hbar} (k_1+k_2)}$;  this last expression is $\geq v_{c,f}^{\mathrm{hab}}$ because $\epsilon_f(k_i)\geq
{\hbar}k_i v_{c,f}^{\mathrm{hab}}$ for all $k_i$.}
\be
v_{c,f}^{L} = \inf_{\kk_1,\kk_2} \frac{\frac{\hbar^2(\kk_1+\kk_2)^2}{2M}+\epsilon_{f,\kk_1}+
\epsilon_{f,\kk_2}}{\hbar|\kk_1+\kk_2|}
\label{eq:vcfun}
\ee
In practice we shall first minimise over $\kk_1$ for a fixed $\qq=\kk_1+\kk_2$, and then minimise over $\qq$. 
We are thus led to the more operational writing, also formally equivalent to the one of equation (\ref{eq:vcL}):
\be
v_{c,f}^{L} = \inf_\qq v_{f,\qq}\ \ \mbox{with}\ \ v_{f,\qq}=\frac{\frac{\hbar^2 q^2}{2M} +\epsilon_{f,\qq}^{\mathrm{eff}}}{\hbar q},
\label{eq:vcfut}
\ee
where $\epsilon_{f,\qq}^{\mathrm{eff}}$ is the lower border of the two fermionic quasi-particle continuum at fixed total wavevector $\qq$:
\be
\epsilon_{f,\qq}^{\mathrm{eff}} \equiv \inf_{\kk_1} \,(\epsilon_{f,\kk_1} + \epsilon_{f,\kk_2=\qq-\kk_1})
\label{eq:def_eps_eff}
\ee
The dispersion relation $\kk\mapsto \epsilon_{f,\kk}$ is a smooth function of the wavevector and diverges at infinity,
so that the infimum in equation (\ref{eq:def_eps_eff}) is reached in a stationary (of zero gradient with respect $\kk_1$) 
point of the function to minimise. Due to the isotropy of the dispersion relation,
\be
\epsilon_{f,\kk} = \epsilon_f(k),
\ee
the gradient is zero if and only if
\be
\epsilon_f'(k_1)\hat{\kk}_1 = \epsilon_f'(k_2)\hat{\kk}_2
\ee
where $\hat{\kk}_i$ is the  direction $\kk_i/k_i$ of the vector $\kk_i$, $\epsilon_f'(k)$ is the derivative of the
function $\epsilon_f(k)$, and where one still has $\kk_2=\qq-\kk_1$. In general this thus leads to four possible branches
of stationarity:
\be
\mbox{(i) : } \kk_1=\kk_2=\qq/2, \hspace{0.5cm} \mbox{(ii) : } \hat{\kk}_1=\hat{\kk}_2, k_1\neq k_2, \hspace{0.5cm}
\mbox{(iii) : } \hat{\kk}_1=-\hat{\kk}_2, \hspace{0.5cm} \mbox{(iv) : } \epsilon_f'(k_1)=\epsilon_f'(k_2)=0
\label{eq:branches}
\ee
In the two intermediate cases, the derivatives $\epsilon_f'(k_1)$ and $\epsilon_f'(k_2)$ are, of course, respectively equal and opposite.

\begin{figure}[htb]
\begin{center}
\includegraphics[height=5cm,clip=]{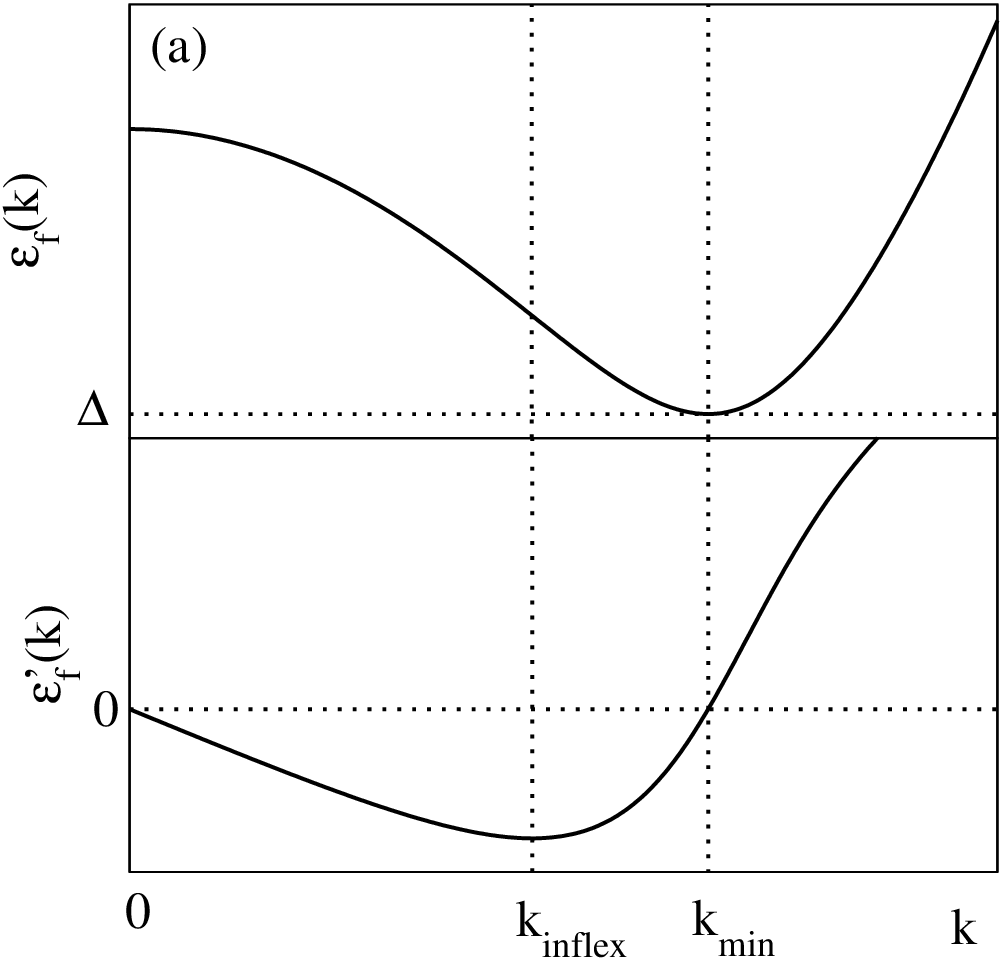} \hspace{2mm}
\includegraphics[height=5cm,clip=]{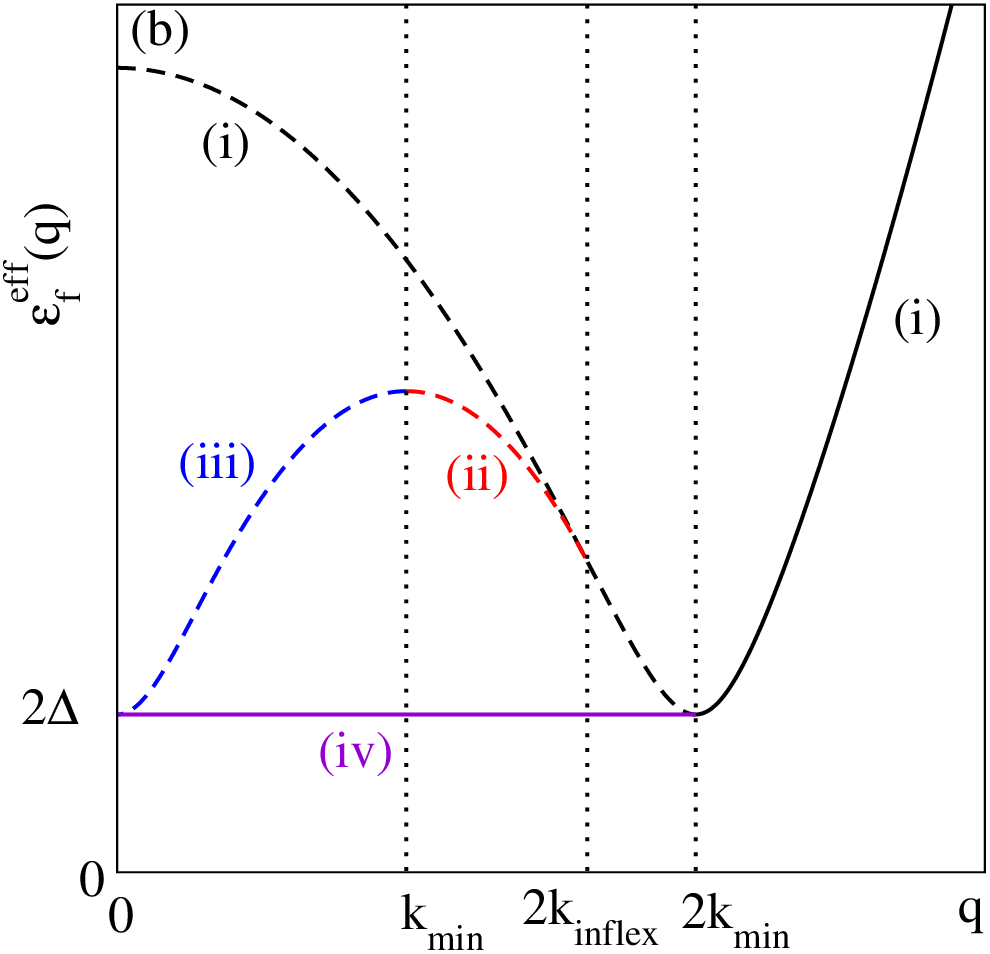} \hspace{2mm}
\includegraphics[height=5cm,clip=]{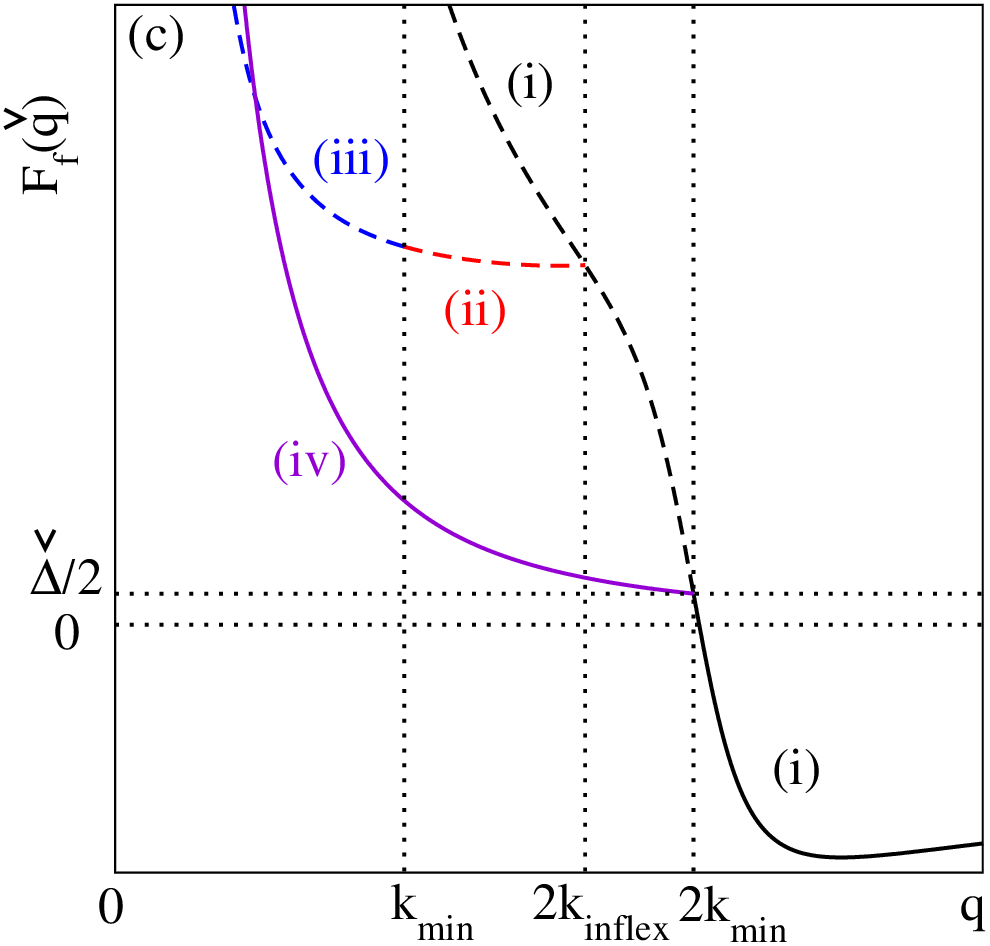}
\end{center}
\caption{For the fermionic excitation branch of BCS theory at positive chemical potential $\mu>0$, (a)
dispersion relation $\epsilon_f(k)$ and its first order derivative $\epsilon_f'(k)$, used in the discussion of the stationarity
branches (\ref{eq:branches}); (b) the stationarity branches as functions of $q$: solid line for the ground branch, dashed line for the other branches;
(c) function $F_{\!f}(\check{q})$ used to minimise $v_{f}(q)$ through a graphical discussion, see equation (\ref{eq:adimf}).
The quantities on the vertical axis of (c) are adimensioned as explained in the text.
\label{fig:ferm}
}
\end{figure}

Let us particularize this discussion to the case of BCS theory, with the dispersion relation
\be
\epsilon_{f,\kk} = \epsilon_f(k) = \left[\left(\frac{\hbar^2 k^2}{2m}-\mu\right)^2+\Delta^2\right]^{1/2}
\label{eq:epsfbcs}
\ee
For a chemical potential $\mu>0$, it exhibits a Mexican hat shape, corresponding to a clear fermionic character, with a 
gap $\Delta$. In this case, $\epsilon_f(k)$ is a concave decreasing function up to the inflexion point $k_{\rm inflex}$,
then it is a convex decreasing function up to its minimum location  $k_{\rm min}$,
\be
k_{\mathrm{min}} = \frac{(2 m\mu)^{1/2}}{\hbar}
\label{eq:kmin}
\ee
and beyond that point it is a convex increasing function. The four branches of stationarity may then be realised.
Contrarily to branch (i), the other branches only exist for low values of $q={|\kk_1+\kk_2|}$. One can fully explore
branch (ii) with  $k_1\in [0,k_{\mathrm{inflex}}]$ and $k_2\in [k_{\mathrm{inflex}},
k_{\mathrm{min}}]$, and one finds that $q=k_1+k_2$ spans $[k_{\mathrm{min}},2k_{\mathrm{inflex}}]$;
similarly, for branch (iii), one can take $0\leq k_1\leq k_{\mathrm{min}}\leq k_2$,  and one finds that
$q=k_2-k_1$ spans $[0,k_{\mathrm{min}}]$.
\footnote{This is due to the fact that $k_1+k_2$ for (ii) and { $k_1-k_2$} for (iii) are increasing functions of $k_1$.
Branches (ii) and (iii) smoothly ($C^\infty$) reconnect in $q=k_{\mathrm{min}}$; to show it, one can
introduce the algebraic quantity $\bar{k}_1\in { [}-k_{\mathrm{min}},
k_{\mathrm{inflex}}]$ and the corresponding extension $\phi(\bar{k}_1)=\epsilon_{f,\bar{k}_1\hat{\kk}_1}$
of $\epsilon_f(k_1)$ to negative arguments. The unique solution $k_2{ \geq }k_{\mathrm{inflex}}$
of $\phi'(k_2)=\phi'(\bar{k}_1)$ then leads to a smooth parameterisation $q=\bar{k}_1+k_2(\bar{k}_1)$
of (ii) plus (iii) as a whole. On the contrary, (i) and (ii), { as well as (iv) et (i),} have only a $C^1$ reconnection:
at $q=2 k_{\mathrm{inflex}}$, { the second order derivative is zero for (i) and is equal to 
$-3[\epsilon_f^{(3)}(k_{\rm inflex})]^2/[2\epsilon_f^{(4)}(k_{\rm inflex})]<0$ for (ii);
at $q=2k_{\rm min}$, the second order derivative is zero for (iv) 
and is equal to $\epsilon_f''(k_{\rm min})/2>0$ for (i).}}
Finally, branch (iv) simply corresponds to $k_1=k_2=k_{\mathrm{min}}$,  and to
$q$ varying from $0$ to $2k_{\mathrm{min}}$. On its existence domain, 
branch (iv) is clearly the minimal energy branch, since the two quasi-particle wavevectors $\kk_1$ and $\kk_2$
are at the bottom of the Mexican hat. Beyond this existence domain,
the branches (ii) and (iii) no longer exist so that the minimal energy is reached on branch (i).
This is illustrated in figure \ref{fig:ferm}. For $\mu>0$, we thus keep
\be
\epsilon_f^{\mathrm{eff}}(q)\underset{\mathrm{branch \ (iv)}}{\stackrel{q\leq 2k_{\rm min}}{=}}2\Delta, \hspace{2cm}
\epsilon_f^{\mathrm{eff}}(q)\underset{\mathrm{branch \ (i)}}{\stackrel{q\geq 2k_{\rm min}}{=}}2\epsilon_f(q/2)
\label{eq:eps_f_eff_syn}
\ee
in agreement with reference \cite{CKS}.
For $\mu<0$, the Cooper pairs of atoms tend to acquire a bosonic character,
and the dispersion relation (\ref{eq:epsfbcs}) is convex,
with a forbidden energy interval of width $(\Delta^2+\mu^2)^{1/2}$. $\epsilon_f(k)$ is then a monotonically increasing
function for $k>0$, and only branch (i) is realized.

To obtain the contribution of the BCS fermionic branch to Landau critical velocity, it remains to minimise
the function $v_{f,\qq}=v_f(q)$ in equation (\ref{eq:vcfut}). This differentiable function diverges at $q=0$ and $q=\infty$, so that
it reaches its minimum with a vanishing derivative at some point $q_0$, $v_f'(q_0)=0$. One switches to dimensionless variables
by taking the wavenumbers $q$ and $q_0$ in units of $(2m|\mu|)^{1/2}/\hbar$ (which is $k_{\mathrm{min}}$ for $\mu>0$),
the energies $\epsilon_f^{\mathrm{eff}}$ and $\Delta$ in units of $|\mu|$ and the velocities $v_f(q)$ and $v_{c,f}^{L}$ in
units of $[|\mu|/(2m)]^{1/2}$, which leads to $\check{v}_f(\check{q})=\alpha\check{q}+\frac{\check{\epsilon}_f^{\rm eff}(\check{q})}{\check{q}}$
and to the elegant implicit equation
\be
\alpha = F_{\!f}(\check{q}_0)\ \ \mbox{with}\ \ F_{\!f}(\check{q}) = \frac{\mathrm{d}}{\mathrm{d}\check{q}} 
\left(-\frac{\check{\epsilon}_f^{\rm eff}(\check{q})}{\check{q}}\right)
\ \ \mbox{and}\ \ \check{v}_{c,f}^{L}= \alpha \check{q}_0 + \frac{\check{\epsilon}_f^{\rm eff}(\check{q}_0)}
{\check{q}_0} 
\label{eq:adimf}
\ee
The check symbol is used to indicate the resulting dimensionless variables, and the mass ratio $\alpha$ is given by equation
(\ref{eq:defalpha}). A graphical discussion of equation  (\ref{eq:adimf}) is readily done. In the non trivial case $\mu>0$,
the various branches of the function $F_{\!f}$, that correspond to the branches (\ref{eq:branches}) of the function $\epsilon_f^{\mathrm{eff}}$,
are shown in figure \ref{fig:ferm}c; only the solid lines, that correspond to minimal energy branches, are relevant.
For $\alpha>\check{\Delta}/2$, the critical velocity is realised on branch (iv), with  $\check{q}_0=(2\check{\Delta}/\alpha)^{1/2}$ and
\be
\check{v}_{c,f}^{L} \underset{\mathrm{branch \ (iv)}}{\stackrel{{ \alpha>\check{\Delta}/2}}{=}} 2(2\check{\Delta} \alpha)^{1/2}
\label{eq:vcf4}
\ee
For $\alpha<\check{\Delta}/2$, it is realised on branch (i), and corresponds to the largest real root of the polynomial
equation of degree four in $\check{v}^2$, that can in principle be expressed in radicals
\footnote{One squares cleverly collected terms in the first and the last equations of (\ref{eq:adimf}), so as to get two polynomial
equations for $\check{q}_0$. $\check{v}$ must be a root of their resultant, and equation (\ref{eq:deg4}) is a divisor of that resultant.}:
\begin{multline}
\label{eq:deg4}
\check{v}^8+8(1-\alpha^2)\check{v}^6 +\check{v}^4 [\check{\Delta}^2(16\alpha^4-80\alpha^2-8)+16\alpha^4-128\alpha^2+16] \\
+\check{v}^2[32\check{\Delta}^2(4\alpha^2-1)(5\alpha^2+1)+128\alpha^2(5\alpha^2-1)] 
-16(4\alpha^2-1)[4\alpha^2(1+\check{\Delta}^2)-\check{\Delta}^2]^2=0
\end{multline}
In the limit $\alpha\to 0$ of an infinite mass impurity, one recovers the first term of the right-hand side of equation (\ref{eq:cks}), that
indeed originates from the fermionic excitation branch  \cite{CKS}.
It remains to determine the nature of the transition on the critical velocity when it passes from branch (iv) to branch (i),
due to a continuous variation of the mass ratio $\alpha$ or of the dimensionless gap $\Check{\Delta}$ (resulting from a variation
of the Fermi gas interaction strength).
As it is apparent on the graphical discussion, $\check{q}_0$ is continuous at the transition, so is $\check{v}_{c,f}^{L}$.
Taking the derivative of the third equation of (\ref{eq:adimf}) with respect to $\alpha$ at fixed $\check{\Delta}$, and
using the first equation of (\ref{eq:adimf}), one finds that the first order derivative of $\check{v}_{c,f}^{L}$ is also continuous:
\be
\frac{\mathrm{d}}{\mathrm{d}\alpha} \check{v}_{c,f}^{L} = \check{q}_0
\label{eq:deriv1}
\ee
The second order derivative of $\check{v}_{c,f}^{L}$ is discontinuous: taking the derivative of (\ref{eq:deriv1}) and of the
first equation of (\ref{eq:adimf}) with respect to $\alpha$, and also taking the derivative of the function $F_{ f}$
on the branches (iv) and (i), one obtains
\be
\left[\frac{\mathrm{d}^2}{\mathrm{d}\alpha^2} \check{v}_{c,f}^{L} \left(\alpha=\frac{\check{\Delta}^+}{2}\right)\right]^{-1} - 
\left[\frac{\mathrm{d}^2}{\mathrm{d}\alpha^2} \check{v}_{c,f}^{L} \left(\alpha=\frac{\check{\Delta}^-}{2}\right)\right]^{-1}=
\frac{1}{4} \check{\epsilon}_f''(\check{k}=1) { = \frac{1}{\check{\Delta}}}
\ee
so that the critical velocity $\check{v}_{c,f}^{L}$ exhibits a {\sl second} order transition on the line $\alpha=\check{\Delta}/2$, see 
figure \ref{fig:domaines}a. In the more straightforward case of a negative chemical potential, $v_{c,f}^{L}$ is always realised on branch (i)
and cannot exhibit any transition.

\section{Critical velocity on the bosonic branch} 
\label{subsec:vcb}

Conservation of the number of fermions of course does not prevent the impurity, when it moves in the superfluid, from creating a single quantum
on the superfluid bosonic excitation branch: this indeed corresponds to excitation of a collective motion of the Cooper pairs of atoms, 
similar to a sound wave. The critical velocity attached to the bosonic branch is then simply given as in equation (\ref{eq:vcL}) by
\be
v_{c,b}^{L} = \inf_{\qq\in\mathcal{D}} v_{b,\qq} \ \ \ \mbox{with} \ \ \ v_{b,\qq}=\frac{\frac{\hbar^2q^2}{2M} + \epsilon_{b,\qq}}{\hbar q}
\label{eq:vcb}
\ee
The bosonic quasi-particle dispersion relation  $\qq\mapsto \epsilon_{b,\qq}=\epsilon_b(q)$ is however more difficult to describe than
the one of the fermionic quasi-particles. Its existence domain $\mathcal{D}$ in the wavevector space, over which one must minimise 
$v_{b,\qq}\!=\!v_b(q)$
in equation (\ref{eq:vcb}), is in itself not so easy to determine. As shown in reference  \cite{CKS}, it is not always
a compact or even connected set. What is very generally known, thanks to superfluid hydrodynamics, is that the branch 
does reach the low wavenumbers $q\to 0$, in a way that is linear with $q$:
\be
\epsilon_{b}(q)\underset{q\to 0}{\sim} \hbar c q,
\label{eq:bson}
\ee
where the coefficient $c$ is simply the sound velocity in the Fermi superfluid, that can be deduced from the gas equation of state 
through the well-known expression $mc^2=\rho \frac{\mathrm{d}\mu}{\mathrm{d}\rho}$. It is also known that $\epsilon_{b}(q)$
must be less than the border $\epsilon_f^{\rm eff}(q)$ of the two fermionic excitation continuum at fixed total wavevector $\qq$.
Otherwise the collective motion of the pairs would damp because its energy, defined as a pole of the dynamic structure factor, 
would become complex \cite{Minguzzi}.

One can obtain $\epsilon_{b}(q)$ numerically at a level of approximation consistent with the BCS theory used in section \ref{subsec:vcf}
thanks to the RPA \cite{Anderson1958}, that was implemented in great details in reference \cite{CKS} not only in the weakly interacting
regime \cite{Minguzzi} but even for arbitrarily strong interactions in the superfluid 
\footnote{\label{note:gacks} In practice, we use dichotomy to solve for $\omega=\epsilon_{b}(q)/\hbar$ the equation $f(\omega,q)=1$,
where, at fixed $q$, $f=I_{11} I_{22}/(\omega^2 I_{12}^2)$ is a decreasing function of $\omega$ that tends to $+\infty$ at  $\omega=0$.
The double integrals $I_{12}$, $I_{11}$ et $I_{22}$ are given by equations (15), (16) and (17) of reference \cite{CKS}.
For all $q\leq 2k_{\mathrm{min}}$, $f(\omega,q)\to -\infty$ logarithmically when $\omega \to \epsilon_f^{\rm eff}(q)^-/\hbar=2\Delta^-/\hbar$,
because $I_{11}\to+\infty$ \cite{CKS} and $I_{22}<0$ $\forall\omega\in [0,2\Delta/\hbar]$, which ensures the existence of a root $\omega \in [0, 2\Delta/\hbar[$.
As a consequence $\mathcal{D}$ contains at least all the wavevectors of modulus $q\leq 2k_{\mathrm{min}}$ \cite{CKS}.}
\footnote{ Similarly to BCS theory, the RPA is only qualitatively correct in the strongly interacting regime, and deviations from the RPA spectrum
can be measured on the collective excitation modes \cite{Altmeyer}.}.
The results on the existence domain are summarized in table \ref{tab:exist};
the interactions are parameterized both by $\Delta/\mu$ and by the more usual quantity $1/(\kf a)$, where $a$ is the $s$-wave scattering length
of opposite spin fermions and $\kf=(3\pi^2\rho)^{1/3}$ is the Fermi wavenumber of the spin $1/2$ unpolarised ideal Fermi gas with the same density
$\rho$ as the superfluid.
It remains to minimise the function $v_b(q)$ over the existence domain $\mathcal{D}$, according to the various forms that it takes.

\begin{table}
\begin{center}
\begin{tabular}{c|c|c|c}
$1/(\kf a)$ & $\Delta/\mu$ & $\mathcal{D}$ & form of $\epsilon_{b}(q)$ \\
\hline
$>0.161$    & $>1.729$ or $<0$ & $[0,+\infty[$ & convex \\
\hline
$\in ]0; 0.161[$  & $\in ]1.162 ; 1.729[$ & $[0,q_{\mathrm{sup}}] \cup [q_{\mathrm{inf}},+\infty[$ & convex if $\Delta/\mu>1.71$\\
\hline
$<0$        & $\in ]0; 1.162[$    & $[0,q_{\rm sup}]$ & concave if $\Delta/\mu< 0.88$\\
\end{tabular}
\end{center}
\caption{Existence domain $\mathcal{D}$ of the superfluid bosonic excitation branch, more precisely ensemble of its accessible wavenumbers $q$,
according to the RPA of reference \cite{CKS}, and (from our own study) convexity of the dispersion relation $q\mapsto \epsilon_b(q)$.
The loss of convexity (of concavity) is due to the emergence of a concave (convex) part at large (low) wavenumbers $q$.
There is always convexity on the component $[q_{\mathrm{inf}},+\infty[$ when it exists.  At $q=q_{\mathrm{sup}}$ and $q=q_{\mathrm{inf}}$,
the bosonic branch meets the two fermionic excitation continuum at the considered total wavenumber, that is $\epsilon_b(q)=\epsilon_f^{\mathrm{eff}}(q)$.
An important result of reference \cite{CKS} is that one always has $q_{\mathrm{sup}}> 2 k_{\mathrm{min}}$ defined in equation (\ref{eq:kmin}).
The parameter $1/(\kf a)$ is more usually used than $\Delta/\mu$ to measure the interaction strength, see text, and it is deduced
here from the BCS equation of state  \cite{CKS}. $1/(\kf a)=0$ is the unitary limit, reached for $\Delta/\mu=1.162\ldots$,
and $\mu<0$ if and only if $1/(\kf a)>0.553\ldots$. Note the notation paradox $q_{\mathrm{sup}} < q_{\mathrm{inf}}$.
\label{tab:exist}}
\end{table}

When the bosonic branch exists at all wavenumber, for example for $\mu<0$, it turns out that the dispersion relation $q\mapsto \epsilon_b(q)$
is convex and always above its tangent at the origin. Then $\epsilon_b(q)\geq \hbar c q$ for all  $q$, the absolute minimum of $v_b(q)$
is reached in $q=0$ and $v_{c,b}^{L}=c$. We go on in the discussion assuming that $\mu>0$.

Minimisation of $v_b(q)$  on the second connected component of $\mathcal{D}$, that is $q\in [q_{\mathrm{inf}},+\infty[$,
when it exists, is also rather simple. One finds that the energy $\epsilon_b(q)$ is extremely close to its ``ceiling" $\epsilon_f^{\mathrm{eff}}(q)$;
since $q_{\mathrm{inf}}$ is always larger than the root of $F_{\!f}(\check{q})$ [see figure \ref{fig:ferm}c],
the three functions $q\mapsto \epsilon_f^{\mathrm{eff}}(q)/q$, $q\mapsto \epsilon_b(q)/q$  and $q\mapsto v_b(q)$
are increasing functions for $q\geq q_{\mathrm{inf}}$. One then numerically checks that the minimum $v_b(q_{\mathrm{inf}})$ of $v_b(q)$
over that second connected component is always larger than the sound velocity, and is thus irrelevant.

It remains to minimise $v_b(q)$ over the existence interval $[0,q_{\mathrm{sup}}]$, and this can in general lead to three different cases:
the absolute minimum is (a) at $q=0$, (b) at $q=q_{\mathrm{sup}}$ or (c) at a point $q_0$ inside the interval.
First one studies the local minima of $v_b(q)$ according to these three cases, then one sorts them.

\begin{figure}[htb]
\begin{center}
\includegraphics[height=5cm,clip=]{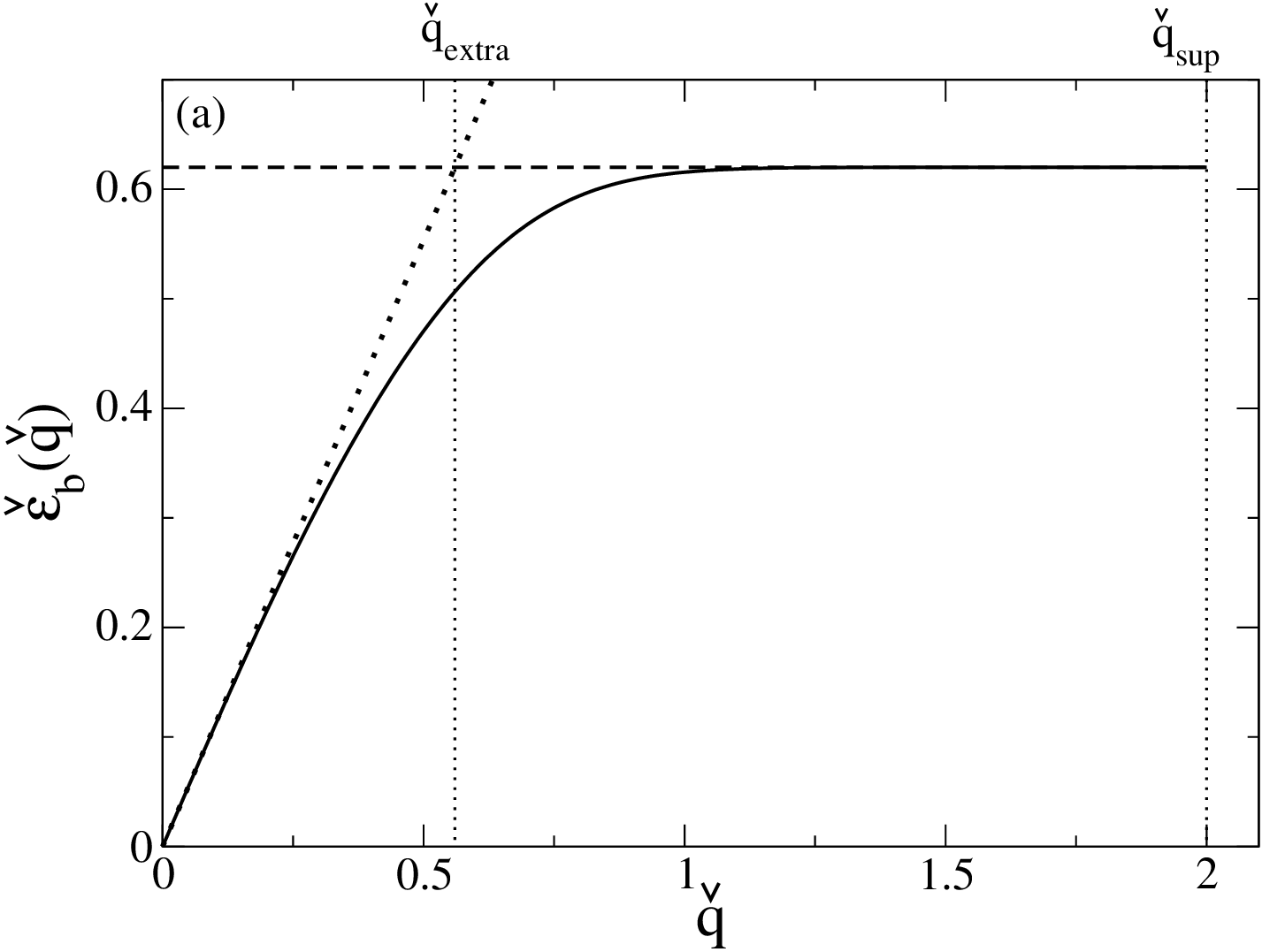} \hspace{5mm}
\includegraphics[height=5cm,clip=]{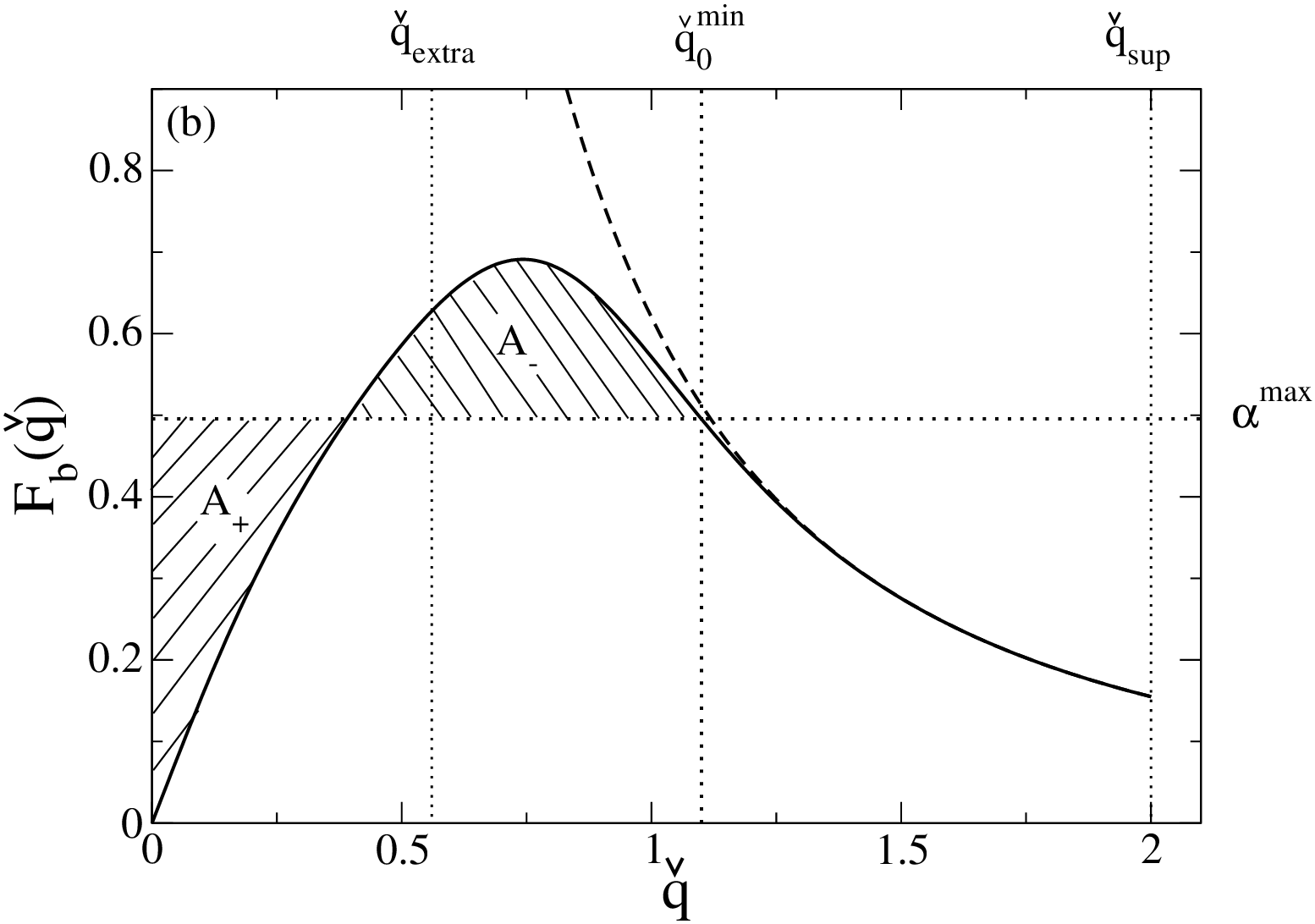}
\end{center}
\caption{For the value $\Delta/\mu=0.31$  taken as an example: (a) dimensionless dispersion relation $\check{\epsilon}_b(\check{q})$ 
of the bosonic branch (solid line), that has a phononic start (dotted line) and is bounded by the border of the two fermionic excitation 
continuum $\check{\epsilon}_f^{\mathrm{eff}}(\check{q})$ at the considered total wavevector (dashed line), and
(b) auxiliary function $F_{\!b}(\check{q})$ for the graphical discussion of the minimisation of $v_b(q)$. Here the bosonic branch only exists
over the compact interval $[0,q_{\mathrm{sup}}]$, and $v_b(q)$ has a local minimum at $q_0$ in the interior of that interval if and only if
the mass ratio $\alpha$ is in between $F_{\!b}(\check{q}_{\mathrm{sup}})$ and $\sup_{\check{q}} F_{\!b}(\check{q})$.
The dashed line in (b) corresponds to the approximation $F_{\!b}(\check{q})\simeq F_{\!f}^{\mathrm{(iv)}}(\check{q})=2\check{\Delta}/\check{q}^2$;
this is a legitimate approximation close enough to $q=q_{\mathrm{sup}}$ and for a small enough value of $\check{\Delta}$. This is indeed
the case here, even for the minimal value $\check{q}_0^{\mathrm{min}}$ of $\check{q}_0$ (that is for the maximal value $\alpha^{\rm max}$
of $\alpha$) that can be accessed in the zone $B_{q_0}$ of figure \ref{fig:domaines}b for a fixed $\check{\Delta}$, such that
$A_+=A_-$ in equation (\ref{eq:dg}).
\label{fig:Fbex}
}
\end{figure}

\noindent{\sl The local minima~:} 
We introduce the same dimensionless variables, auxiliary function $F(q)$ and graphical discussion as for the fermionic branch:
\be
\check{v}_{b}(\check{q}) = \alpha \check{q} + \frac{\check{\epsilon}_b(\check{q})}{\check{q}} \ \mbox{and}\ \ 
F_{\!b}(\check{q})=\frac{\mathrm{d}}{\mathrm{d}\check{q}}
\left(-\frac{\check{\epsilon}_b(\check{q})}{\check{q}}\right), \ \ \mbox{so that}\ \ \
\frac{\mathrm{d}}{\mathrm{d}\check{q}} \check{v}_{b}(\check{q}) = \alpha-F_{\!b}(\check{q})
\label{eq:outils}
\ee
A first remark is that $F_{\!b}(0)=0$ {and $\frac{\mathrm{d}}{\mathrm{d}\check{q}} \check{v}_{b}(0)=\alpha>0$,}
so that $v_b(q)$ always has a minimum at $q=0$, because the first correction to the linear term in equation (\ref{eq:bson}) is cubic
(the Taylor expansion of $[\epsilon_b(q)]^2$ according to the RPA only contains even powers of $q$).
A second remark is that the function $v_b(q)$ has a minimum at $q=q_{\mathrm{sup}}$ if $\frac{\mathrm{d}}{\mathrm{d}\check{q}} \check{v}_{b}(\check{q}_{\rm sup})<0$, that is if $\alpha<F_{\!b}(\check{q}_{\mathrm{sup}})$.
A last remark is that $v_b(q)$ has a local minimum at $q_0\in ]0,q_{\mathrm{sup}}[$ if its first order derivative vanishes at
$q_0$ and if its second order derivative is positive. Graphically this means that $\check{q}\mapsto F_{\!b}(\check{q})$ crosses
the horizontal line of ordinate $\alpha$ inside the interval with a negative derivative, that is
from {\sl top to bottom}. This may be realised for some value of $\alpha$ if and only if the continuous function $F_{\!b}(\check{q})$ 
has a strictly positive maximum in $]0,q_{\mathrm{sup}}[$, as in figure \ref{fig:Fbex}.

\noindent{\sl The global minimum $v_{c,b}^{L}$:} The values $v_b(0)=c$ and $v_b(q_{\mathrm{sup}})$ can be directly compared,
after a numerical calculation of $c$ and $q_{\mathrm{sup}}$, since $\epsilon_b(q)$ and the analytically known $\epsilon_f^{\mathrm{eff}}(q)$
coincide at $q_{\mathrm{sup}}$. When it exists, the local minimum of $v_b(q)$ at $q_0\in ]0,q_{\mathrm{sup}}[$ is actually
smaller than $v_b(q_{\mathrm{sup}})$, since $F_{\!b}(\check{q})$ remains below the horizontal line of ordinate $\alpha$ over the
interval $[\check{q}_0,\check{q}_{\mathrm{sup}}]$ so that $v_b(q)$ is an increasing function over that interval.
It can also easily be compared to the sound velocity: after integration of the third equation of (\ref{eq:outils}), one finds that
\be
\check{v}_b(\check{q}_0)-\check{c} = \int_0^{\check{q}_0} \mathrm{d}\check{q}\, [\alpha-F_{\!b}(\check{q})] = A_+ - A_-
\label{eq:dg}
\ee
where $A_+$ and $A_-$ are the (positive) areas of the zones delimited by the graph of $F_{\!b}(\check{q})$ and by the horizontal straight line
of ordinate $\alpha$, respectively below and above that straight line, for $\check{q}$ spanning $[0,\check{q}_0]$. They are the hatched zones
in figure \ref{fig:Fbex}b, plotted in the particular case $A_+=A_-$, that is for the value $q_0^{\mathrm{min}}$ of $q_0$ below which
${v}_b({q}_0)$ ceases to be strictly less than $c$.

The result of the global minimisation is shown in figure \ref{fig:domaines}b. The boundary between the zones $v_{c,b}^{L}=v_b(q_{\mathrm{sup}})$
and $v_{c,b}^{L}=v_b(q_0)$ corresponds to the limiting case $q_0\to q_{\mathrm{sup}}$, that is to the equation $\alpha=F_{\!b}(\check{q}_{\mathrm{sup}})$;
it leads, as shown by generalisation of the property (\ref{eq:deriv1}) of $v_{c,f}^{L}$ [see (\ref{eq:diffv_alpha}) and footnote \ref{note:tauto}],
to a second order transition for $v_{c,b}^{L}$, that is 
to a discontinuity of its second order derivative in the direction orthogonal to the boundary. The other boundaries, where
$c=v_b(q_0)$ or $c=v_b(q_{\rm sup})$, lead to first order transitions for $v_{c,b}^{L}$, that is to a discontinuity of its first order derivative,
since the location of the minimum of $v_b(q)$
jumps from $0$ to $q_0^{{\mathrm{min}}}>0$ or $q_{\mathrm{sup}}$. Note the existence of a triple point at the confluence of the three zones.

\section{Synthesis: global critical velocity of the particle}
\label{subsec:vc}

The global Landau critical velocity  for a moving particle in the superfluid is given by the smallest of the two velocities
$v_{c,f}^{L}$ and $v_{c,b}^{L}$ of the previous sections. For $\mu<0$ or $\Delta/\mu>1.729$, one always has $v_{c,b}^{L}=c<v_{c,f}^{L}$,
so that $v_{c}^{L}$ is identically equal to the sound velocity and has a bosonic origin. 
As shown in table \ref{tab:exist} indeed, the bosonic excitation branch then exists for all $q$ with an energy $\epsilon_b(q)$
everywhere $<\epsilon_f^{\rm eff}(q)$, so that $v_b(q)<v_f(q)$ and $v_{c,b}^L<v_{c,f}^L$ in equations (\ref{eq:vcfut}) and (\ref{eq:vcb});
furthermore, the convexity of $q\mapsto \epsilon_b(q)$ implies $v_{c,b}^{L}=c$, see section \ref{subsec:vcb}.
For $\mu>0$ and $\Delta/\mu<1.729$, the diagram in the plane $(\Delta/\mu,\alpha=m/M)$ in figure \ref{fig:diagvc}a shows that 
the critical velocity has a fermionic origin ($v_{c,f}^{L}<v_{c,b}^{L}$) in a sort of triangle with one curved side and with a basis
lying on the $\alpha=0$ axis where the impurity has an infinite mass; its maximal extension on that axis agrees with the crossing point
$\Delta/\mu\simeq 1.38$ of the two terms in the right-hand side of (\ref{eq:cks}). Everywhere else, the critical velocity has a
bosonic origin. One can explain in simple terms why this property necessarily holds to the right of this crossing point:
$v_f(q)$ and thus $v_{c,f}^{L}$ in equation (\ref{eq:vcfut}) are increasing functions of $1/M$, and the corresponding dimensionless
quantities are increasing functions of $\alpha$; according to (\ref{eq:cks}), $v_{c,f}^{L}(\alpha=0)>c$ as soon as $\Delta/\mu> 1.38$,
so that one also has $v_{c,f}^{L}(\alpha)>c$ for all $\alpha>0$.

\begin{figure}[H]
\begin{center}
\includegraphics[height=5.5cm,clip=]{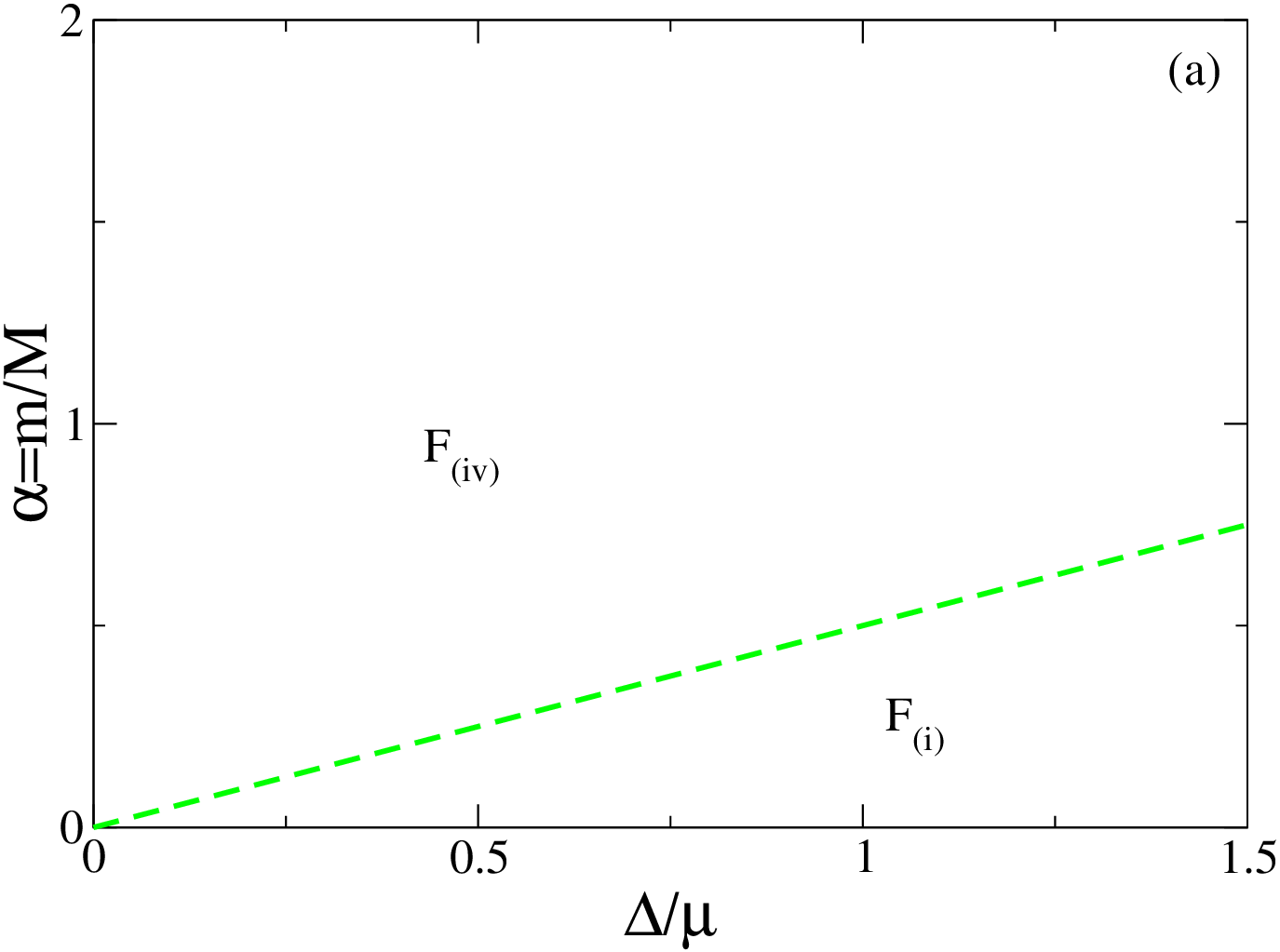} \hspace{5mm}
\includegraphics[height=5.5cm,clip=]{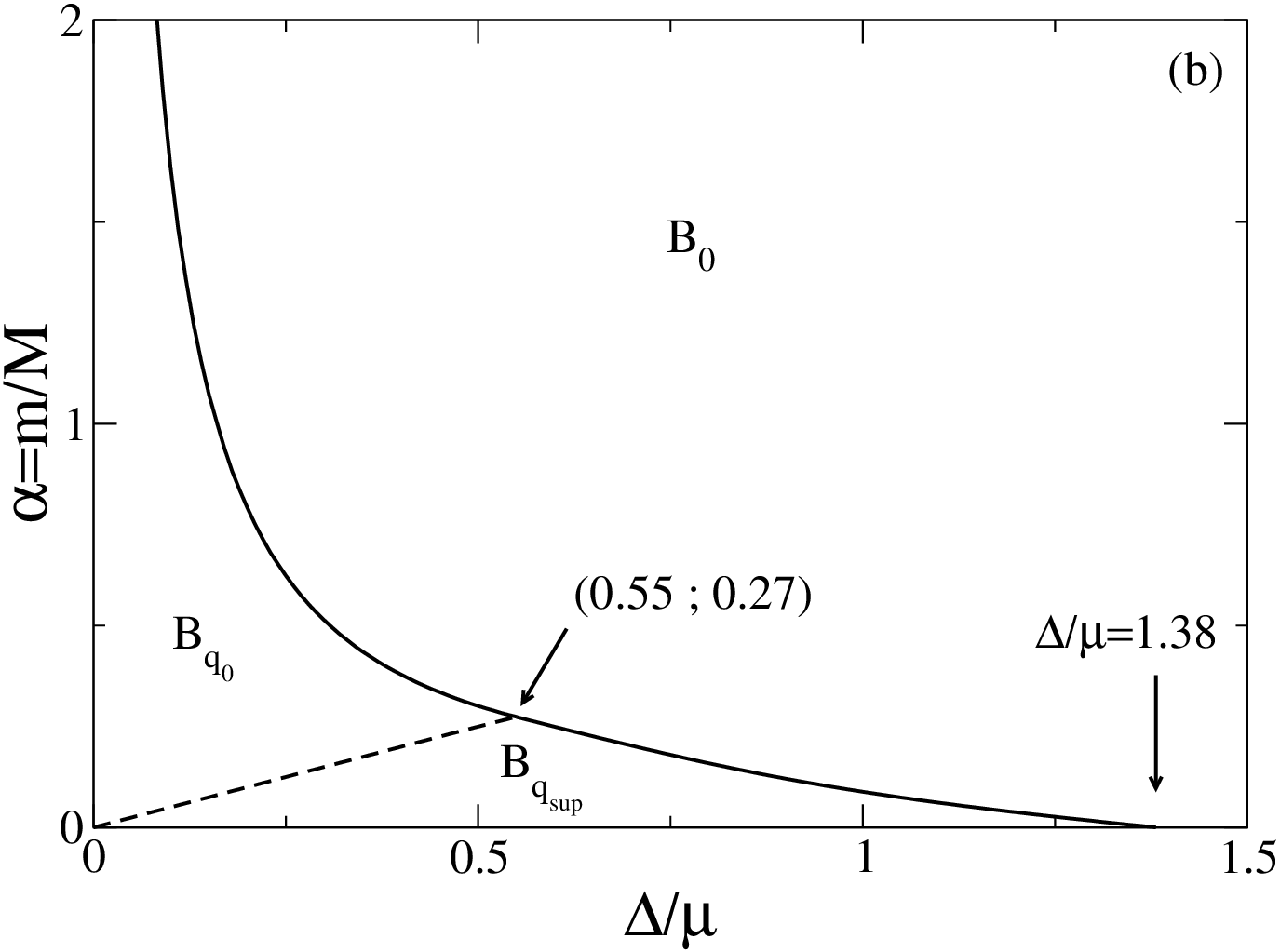} 
\end{center}
\caption{Diagram in the plane ($\check{\Delta}=\Delta/\mu,\alpha=m/M$), $\mu>0$, indicating (a) for the fermionic excitation branch of the
superfluid, on which stationarity branch of equation (\ref{eq:branches}) the critical velocity $v_{c,f}^{L}$ is realised (with the 
self-explanatory notations $F_{\mathrm{(i)}}$ and $F_{\mathrm{(iv)}}$), and (b) for the bosonic excitation branch, if the 
critical velocity $v_{c,b}^{L}$
is reached at the lower endpoint $q=0$ (zone $B_0$), at the upper endpoint $q=q_{\mathrm{sup}}$ (zone $B_{q_\mathrm{sup}}$) or in the
interior of its existence interval, $q=q_0\in]0,q_{\mathrm{sup}}[$ (zone $B_{q_0}$).
The solid (dashed) lines indicate a first order (second order) transition for $v_{c,f}^{L}$ or $v_{c,b}^{L}$, that
is with a discontinuous first order (second order) differential.
\label{fig:domaines}
}
\end{figure}

\begin{figure}[H]
\begin{center}
\includegraphics[height=5.5cm,clip=]{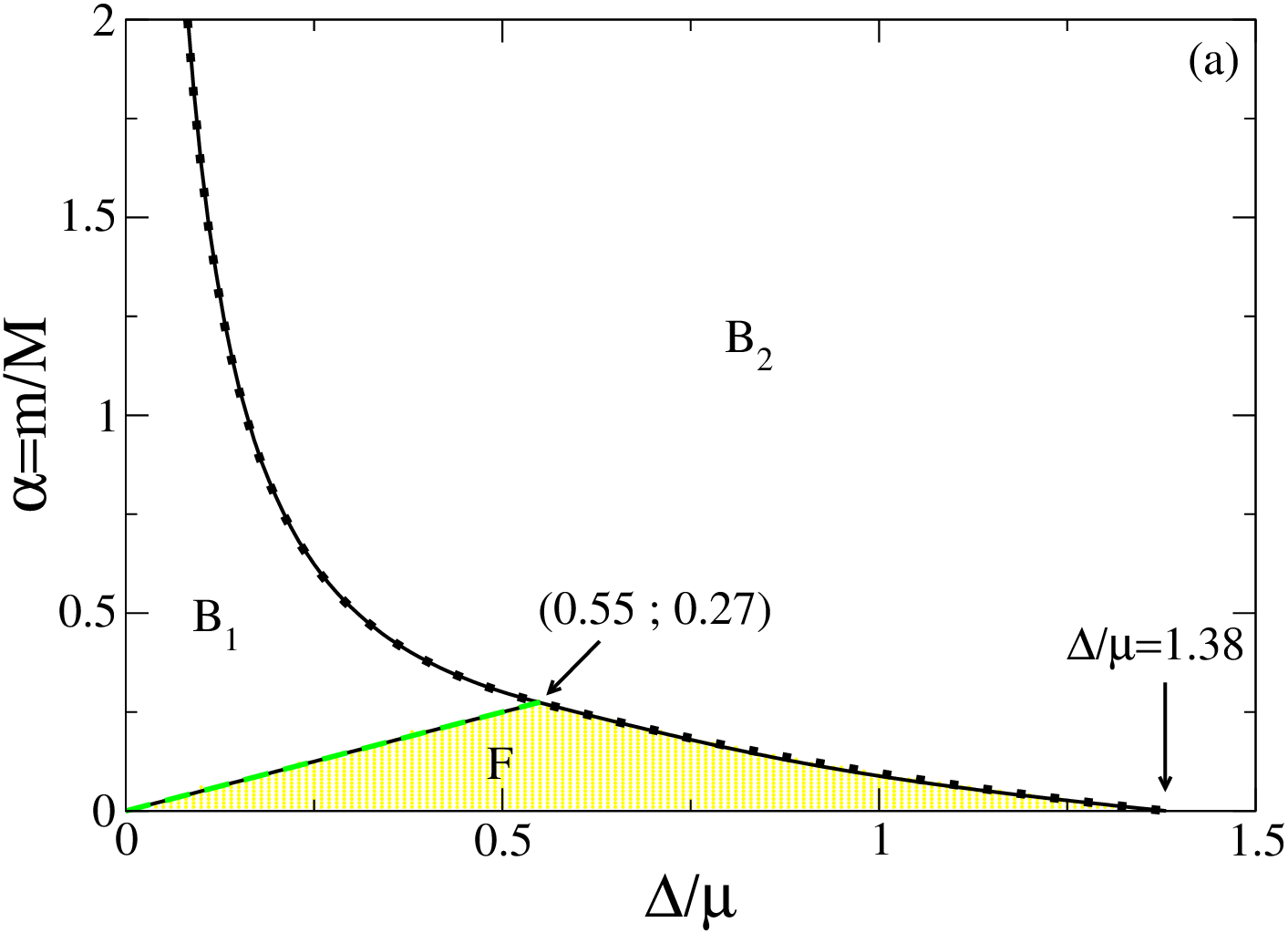} \hspace{5mm}
\includegraphics[height=5.5cm,clip=]{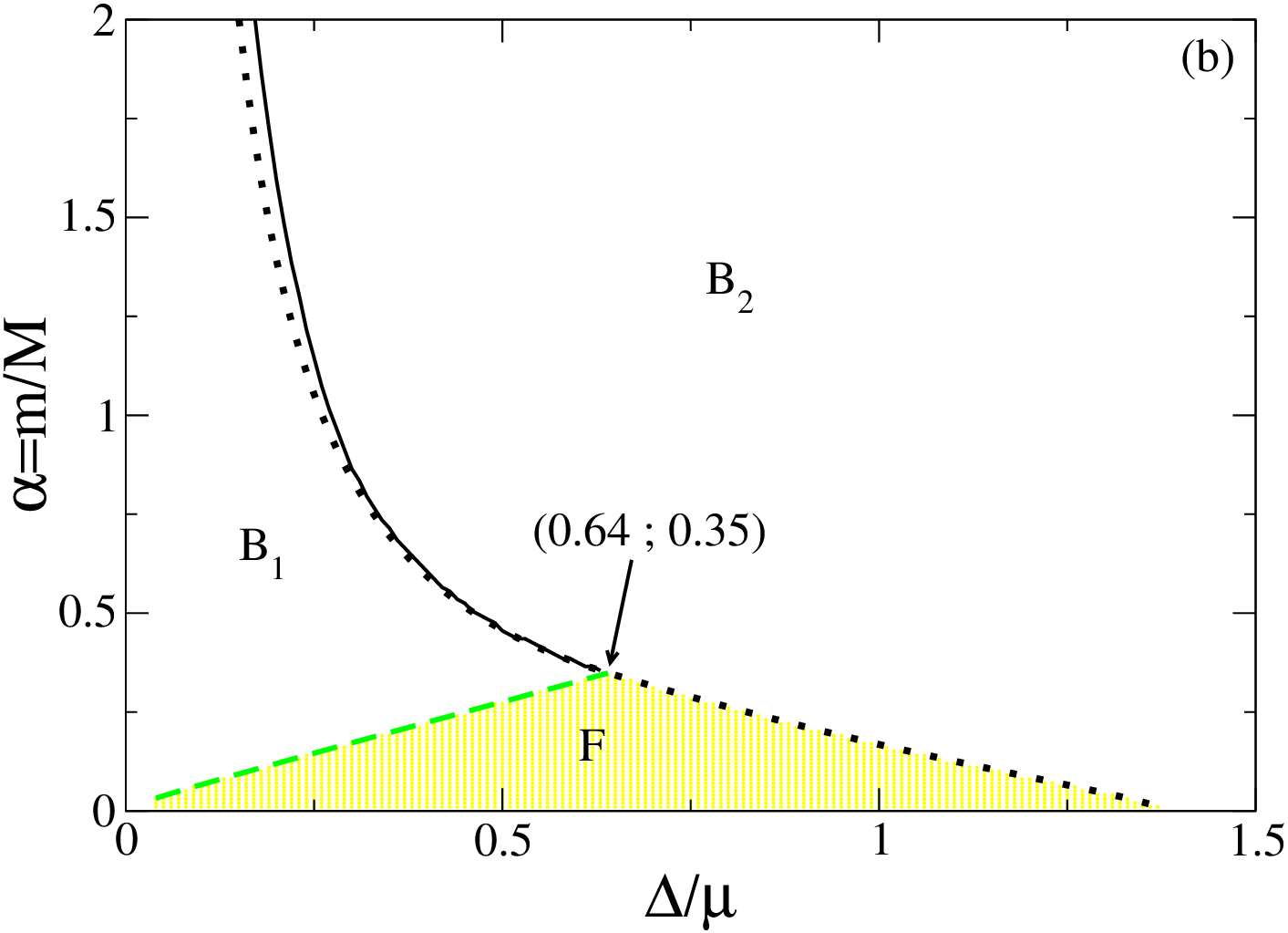}
\end{center}
\caption{(a)  Diagram in the plane $(\check{\Delta}=\Delta/\mu,\alpha=m/M)$, $\mu>0$, showing in which zone the global critical velocity
$v_c^{L}$ of the impurity of mass $M$ has a bosonic origin (case $v_{c,b}^{L}<v_{c,f}^{L}$, indicated by the letter B and a uniform
white background) or a fermionic origin (case $v_{c,f}^{L}<v_{c,b}^{L}$, indicated by the letter F and a yellow hatched background).
The bosonic domain is split in two sub-domains $B_1$ and $B_2$ by the first order transition line between the zones $B_{q_{\mathrm{sup}}}$
and $B_{q_0}$ [black solid line with abscissas $\check{\Delta}<0.55$] of figure \ref{fig:domaines}b.
Unexpectedly, the $B_2-F$ boundary is close to the first order transition line between the zones $B_{q_{\mathrm{sup}}}$
and $B_0$ [black solid line with abscissas $\check{\Delta}>0.55$] of figure \ref{fig:domaines}b.
Remarkably also, the unmasked part of the second order transition line for $v_{c,f}^{L}$
[green dashed line] and the second order transition line between the zones $B_{q_0}$ and $B_{q_{\mathrm{sup}}}$
[black dashed line], that bracket the $B_1-F$ boundary, are in practice indistinguishable;
an analytical study shows that the $B_1-F$ and $B_{q_0}-B_{q_{\mathrm{sup}}}$ boundaries actually exactly coincide,
and that $v_b(q)$ and $v_f(q)$ on that boundary have their minimum in $q=q_{\rm sup}$, and that $v_c^{L}$ exhibits
a second order transition.
Last, the line $v_{c,f}^{L}=c$ [black dotted line] reproduces exactly  the $B_2-F$ boundary (as it should be) and also 
quite well the $B_2-B_1$ boundary.
(b) Generalisation of the previous diagram to the case of a superfluid of bosonic impurities moving in the Fermi superfluid,
for a fixed ratio  $\mu_B/E_F=0.1$ of the rest frame Bose chemical potential and the Fermi energy of the fermions.
The new critical velocity $v_{c}^{L}$ also exhibits a partition in three zones, separated as in the previous case 
by a discontinuity of its second order ($F-B_1$ boundary) or first order ($F-B_2$ and $B_1-B_2$ boundaries) differential:
the zone $F$ (yellow hatched) where $v_c^{L}=v_{c,f}^{L}$, and the zones $B_1$ and $B_2$, where $v_c^{L}=v_{c,b}^{L}$.
The $F-B_1$ boundary, where $v_b(q)$ and $v_f(q)$ have their minimum at $q=q_{\rm sup}$,
is very close to the $F_{\rm (i)}-F_{\rm (iv)}$ boundary at the considered value of $\mu_B$ 
[green dashed line, interrupted as it reaches zone $B_2$]. The $F-B_2$ boundary is given exactly (as it should be)
by the black dotted line $v_{c,f}^{L}=c+c_B$, where $c$ ($c_B$) is the son velocity in the Fermi (Bose) superfluid at rest,
to the right of the triple point. The $B_1-B_2$ boundary, which is simply the $B_{q_0}-B_0$ boundary at the considered
value of $\mu_B$ [black solid line] on the contrary deviates from the dotted line away from and to the left of the triple point.
\label{fig:diagvc}
}
\end{figure}

It remains to be seen to which extent the transition line(s) predicted for $v_{c,f}^{L}$ and $v_{c,b}^{L}$ [see figure \ref{fig:domaines}]
subsist on the global critical velocity $v_{c}^{L}$, or on the contrary are masked because the critical velocity from the competing 
excitation branch is smaller. We have plotted the transition line for $v_{c,f}^{L}$ [among the stationarity branches (i) and (iv)],
$\alpha=\check{\Delta}/2$, as a dashed green line in figure \ref{fig:diagvc}a. The portion corresponding to $\check{\Delta}>0.55$
is entirely masked by the bosonic critical velocity and is omitted; remarkably, and may be unexpectedly, the portion corresponding to
$\check{\Delta}<0.55$ cannot be distinguished, at the scale of the figure, from the boundary between the bosonic domain and the 
fermionic domain! Similarly, we have plotted the transition lines of $v_{c,b}^{L}$ [depending on the location of the minimum of $v_b(q)$
at $q=0$, $q=q_{\mathrm{sup}}$ or in between] as a black solid (dashed) line for a first (second) order transition.
Two additional remarkable facts arise. First, the black dashed line is in practice indistinguishable from the green dashed line,
and therefore from the boundary between the bosonic and fermionic domains.
Second, the portion of solid line with $\check{\Delta}>0.55$ seems to coincide quite well with another piece of that boundary.
Last, the portion of solid line with abscissas $\check{\Delta}<0.55$ is immersed in the bosonic domain, and splits it in
two subdomains $B_1$ and $B_2$ separated by a first order transition for $v_{c}^{L}$. We shall now give some simple facts
allowing one to understand part of those observations.

\noindent{\sl Some zones of predictable origin:}
The domain $B_{q_{\mathrm{sup}}}$ such that $v_{c,b}^{L}=v_b(q_{\mathrm{sup}})$ necessary corresponds to $v_{c,b}^{L}\geq v_{c,f}^{L}$
so it is, in the final diagram for  $v_{c}^{L}$, entirely masked by the critical velocity induced by the fermionic excitation branch.
At $q=q_{\mathrm{sup}}$, the bosonic excitation branch indeed meets the two fermionic excitation ``ceiling" $\epsilon_{f}^{\mathrm{eff}}(q)$
so that $v_b(q_{\mathrm{sup}})=v_f(q_{\mathrm{sup}}) \geq \inf_q v_f(q)=v_{c,f}^{L}$.
As a consequence, the $B_1-F$ boundary is {\sl above} the transition line between the zones $B_{q_0}$ [where $v_{c,b}^{L}=v_b(q_0)$]
and $B_{q_{\mathrm{sup}}}$ [where $v_{c,b}^{L}=v_b(q_{\mathrm{sup}})$], that is above the black dashed line in figure \ref{fig:diagvc}a.

In a symmetric manner, the domain $F_{\mathrm{(iv)}}$ of  $v_{c,f}^{L}$ corresponding to the stationarity branch (iv),
that is to $\alpha>\check{\Delta}/2$, is entirely masked by the contribution of the bosonic excitation branch.
On branch (iv), indeed, $v_{f}(q)$ reaches its minimum over the interval $q\in [0,2 k_{\rm min}]$,
see figure~\ref{fig:ferm}c. Over this interval, the bosonic excitation branch does exist, since $q_{{\mathrm{sup}}}>2k_{\mathrm{min}}$
as shown by reference  \cite{CKS}, and leads to a velocity $v_b(q)$ everywhere smaller than the velocity $v_f(q)$,
since one has everywhere $\epsilon_b(q) \leq \epsilon_{f}^{\mathrm{eff}}(q)$. As a consequence, $v_{c,b}^{L}$ 
is less than $v_{c,f}^{\rm(iv)}=\inf_{q\in [0,2 k_{\rm min}]} v_f(q)$, and the boundary between $B_1$ and $F$ is
{\sl below} the $F_{\mathrm{(iv)}}-F_{\mathrm{(i)}}$ transition line for $v_{c,f}^{L}$, 
that is below the green dashed line in figure \ref{fig:diagvc}a.

To summarise, the $B_1-F$ boundary is bracketed by the green dashed ceiling $\alpha=F_{\!f}(2)=\check{\Delta}/2$
(see section \ref{subsec:vcf}), and the black dashed floor $\alpha=\lim_{\check{q}_0\to \check{q}_{\mathrm{sup}}}
F_{\!b}(\check{q}_0)=F_{\!b}(\check{q}_{\mathrm{sup}})$ (see section \ref{subsec:vcb}).
But, as the numerics show, the quantity $F_{\!b}(\check{q}_{\mathrm{sup}})$ considered as a function of $\check{\Delta}$
is extremely close to $\check{\Delta}/2$ up to $\check{\Delta}\approx 0.6$ (beyond that value, it starts bending down).
More precisely, for $\check{\Delta}<0.55$, the deviation is less than four per mil and, also, $q_{\mathrm{sup}}$ 
differs from $2k_{\mathrm{min}}$ by less than one per mil. This is why the  $B_1-F$ boundary, the green dashed line
and the black dashed line almost coincide in figure \ref{fig:diagvc}a. We shall be more precise at the end of this section:
we shall show analytically that the $B_1-F$ boundary lies exactly on top of the black dashed line and that two zones of the
plane $(\check{\Delta},\alpha)$ exactly coincide:
\be
B_{q_{0}} { =} B_1
\label{eq:egalite_zones}
\ee

\noindent{\sl Explaining why the dashed lines almost coincide at the $B_1-F$ boundary:}
In the limit  $\check{\Delta}=\Delta/\mu\ll 1$, one can physically understand why the green and black dashed lines almost coincide:
The low $q$ linear part of $\epsilon_b(q)$, when linearly extrapolated, reaches the two fermionic excitation 
ceiling $\epsilon_f^{\mathrm{eff}}(q)\simeq 2\Delta$ at a point $q_{\mathrm{extra}}\approx 2\Delta/(\hbar c)\ll 2 k_{\mathrm{min}}$
since $c\simeq (2\mu/3m)^{1/2}$ in that limit. The bosonic excitation branch actually bends at $q=q_{\mathrm{extra}}$
then closely follows the ceiling until it reaches it at the point $q_{\mathrm{sup}}$, which is very close to $2 k_{\mathrm{min}}$
($\check{q}_{\rm sup}\simeq 2$). As a consequence, the functions $\epsilon_b(q)$ and $\epsilon_f^{\mathrm{eff}}(q)$,
or the functions $F_{\!f}(\check{q})$ and $F_{\!b}(\check{q})$ may be in practice identified on a broad neighbourhood of
 $q_{\mathrm{sup}}$, or of $\check{q}_{\mathrm{sup}}$ far enough to the right of $q_{\rm extra}$ or of $\check{q}_{\rm extra}=3^{1/2}\check{\Delta}$; 
at this stage, one can also assimilate the function $F_{\!f}(\check{q})$ to its expression $\frac{\mathrm{d}}{\mathrm{d}\check{q}} (-2\check{\Delta}/\check{q})=2\check{\Delta}/\check{q}^2$ on the stationarity branch (iv), see equations (\ref{eq:eps_f_eff_syn}) and (\ref{eq:adimf}).
In short:
\be
F_{\!b}(\check{q})\simeq \frac{2\check{\Delta}}{\check{q}^2} \ \ \mbox{for}\ \ \check{\Delta} \ll \check{q}\ \ \mbox{and}\ \ 
 \check{\Delta} \ll 1
\label{eq:approxFb}
\ee
These ideas are successfully illustrated in figure \ref{fig:Fbex}. This explains why  $F_{\!b}(\check{q}_{\mathrm{sup}})
\simeq \check{\Delta}/2$ at low $\check{\Delta}$. What is fortunate here is that $\check{\Delta}=0.55$ is already small enough
from that perspective.

\noindent{\sl $v_{c,f}^{L}$ and $v_{c,b}^{L}$ almost coincide in the zone $B_1$~:}
One may wonder if the previous approximation (\ref{eq:approxFb}) applies not only to $\check{q}=\check{q}_{\mathrm{sup}}$
but also to the location $q_0$ of the absolute minimum of $v_b(q)$, within the zone $B_{q_0}$ of figure \ref{fig:domaines}b.
For a fixed $\check{\Delta}$, this suffices to be checked for the minimal reachable value $q_0^{\mathrm{min}}$ of $q_0$,
corresponding to the maximal value of $\alpha$ reached in that zone, that satisfies $A_+=A_-$ in equation (\ref{eq:dg}).
This is confirmed by the numerics, which means that $\check{q}_0^{\mathrm{min}}$ is always in practice far enough
to the right of the location of the maximum of $F_{\!b}(\check{q})$, since that maximum cannot be predicted by (\ref{eq:approxFb}).
As a remarkable consequence, we obtain that in the zone $B_{q_0}$, that is in the zone $B_1$:
\be
\forall (\check{\Delta},\alpha)\in B_1, v_{c}^{L}=v_{c,b}^{L}\simeq v_{c,f}^{L} 
\label{eq:vcb_est_vcf}
\ee
where equation (\ref{eq:vcf4}) can be used to get $v_{c,f}^{L}$.

\noindent{\sl Line $v_{c,f}^{L}=c$ is remarkable~:}
Now that we have understood the bosonic-fermionic nature of zone $B_1$, that is the validity of the approximation (\ref{eq:vcb_est_vcf}),
we can approximate in a simple way the location of the $B_1-B_2$ boundary, that is of the $B_{q_0}-B_0$ boundary, where the zone $B_0$
is such that $v_{c,b}^{L}=c$. One just needs to solve the equation
\be
v_{c,f}^{L}=c
\label{eq:vcfegalec}
\ee
The corresponding black dotted line in figure \ref{fig:diagvc}a is indeed very close to the black solid line for $\check{\Delta}<0.55$.
Even better, it reproduces {\sl exactly} the $B_2-F$ boundary for $\check{\Delta}>0.55$, where $\alpha<\check{\Delta}/2$ 
and $v_{c,f}^{L}$ now originates from the stationarity branch (i).  One has indeed $v_{b}(q_{\mathrm{sup}})=v_{f}(q_{\mathrm{sup}})\geq v_{c,f}^{L}$
as we have already seen, so that the zone $B_{q_{\mathrm{sup}}}$ cannot compete and the $F-B_2$ transition is a
$F-B_0$ transition given by equation (\ref{eq:vcfegalec}).

\newpage


\noindent{\sl A quasi-coincidence at the $B_2-F$ boundary:}
What ultimately remains to be explained is the quasi-coincidence of the $F-B_2$ and $B_{q_{\mathrm{sup}}}-B_0$ boundaries,
that is of the black dotted line and the black solid line for $\check{\Delta}>0.55$ in figure~\ref{fig:diagvc}a.
This quasi-coincidence is however more approximate than the previous ones, not to say accidental.
It turns out that at the point of the $F-B_2$ boundary with abscissa $\check{\Delta}=0.55$,
$q_{\mathrm{sup}}$ is very close to the location $q=2k_{\mathrm{min}}$ of the minimum of $v_f(q)$;
also, at the endpoint of that boundary with abscissa $\check{\Delta}\simeq 1.38$, $\check{q}_{\mathrm{sup}}\simeq 2.59$
is very close to the location $\check{q}\simeq 2.61$ of the minimum of $\check{v}_f(\check{q})$, which is unexplained.
As a consequence, the $F-B_2$ and $B_{q_{\mathrm{sup}}}-B_0$ boundaries almost touch in their endpoints.
In the intermediate region $0.55 < \check{\Delta}< 1.38$, however, they appreciably deviate in figure \ref{fig:diagvc}a;
numerics confirm that $q_{\mathrm{sup}}$ can significantly deviate from the location of the minimum of $v_f(q)$,
at least by 5\%, but this only leads to a weak deviation of $v_b(q_{\mathrm{sup}})=v_f(q_{\mathrm{sup}})$ from $\inf_q v_f(q)$
because $v_f(q)$ varies only quadratically around its minimum.

\noindent{\sl Order of the transitions and summary:}
We finally give the minimal order of the differentials of $v_c^{L}$ that are discontinuous at the boundaries between the zones
$B_1$, $B_2$ and $F$. Since no boundary has any vertical portion in the plane $(\check{\Delta},\alpha)$, we can limit ourselves
to the derivatives of $v_c^{L}$ with respect to $\alpha$ at fixed $\check{\Delta}$, taking advantage within each zone
of the general property:
\be
\frac{\mathrm{d}}{\mathrm{d}\alpha} v_c^{L} = q_0^{\rm abs}
\label{eq:diffv_alpha}
\ee
where $q_0^{\rm abs}$, a function of $\alpha$, is the location of the absolute minimum $q_{0,b}$ of $q\mapsto v_b(q)$
or $q_{0,f}$ of $q\mapsto v_f(q)$, depending on whether $v_{c}^{L}$ originates from the bosonic or fermionic excitation branch\footnote{
\label{note:tauto}
When $v_{c}^{L}=v_{c,f}^{L}$, this is a tautology of equation (\ref{eq:deriv1}). When $v_{c}^{L}=v_{c,b}^{L}$, one has {\sl either} 
$0<q_{0,b}<q_{\rm sup}$,
in which case $\alpha=F_{\!b}(\check{q}_{0,b})$ and one simply takes the derivative of $\check{v}_b(\check{q}_{0,b}(\alpha))$ with respect to $\alpha$
in equation (\ref{eq:outils}), {\sl or } $q_{0,b}=0$ or $q_{\rm sup}$, in which case $\check{q}_{0,b}$ is locally constant and the result is trivial.}.

Another remarkable, may be even surprising property is that the bosonic excitation branch $q\mapsto \epsilon_b(q)$
is exactly {\sl tangential} to the two fermionic excitation ceiling $q\mapsto \epsilon_f^{\rm eff}(q)$ at the point
of abscissa $q=q_{\rm sup}$ where they meet\footnote{
We use the footnote \ref{note:gacks} and the explicit expressions of the integrals $I_{11}$, $I_{12}$ and $I_{22}$ of reference \cite{CKS}.
Taking the derivative of the implicit equation $f(\omega_b(q),q)=1$ with respect to  $q$, we obtain 
$\frac{\mathrm{d}}{\mathrm{d}q}\omega_b(q)=-\partial_q f/\partial_\omega f$. Taking the partial derivatives $\partial_\omega$ and $\partial_q$
of each integral under the integral sign, we get in the integrand a factor $1/[\epsilon_{f,\kk+\qq/2}+\epsilon_{f,\kk-\qq/2}-\epsilon_b(q)]^2$,
whose tridimensional integral over $\kk$ is infrared divergent when $q\to q_{\rm sup}^-$, that is when $\epsilon_f^{\rm eff}(q)-\epsilon_b(q)\to 0^+$,
since $\check{\epsilon}_{f,\kk+\qq/2}+\check{\epsilon}_{f,\kk-\qq/2}\underset{\kk\to \mathbf{0}}{=}\check{\epsilon}_f^{\rm eff}(\check{q})
+\check{k}^2 (\check{q}^2-4)/\check{\epsilon}_f^{\rm eff}(\check{q})
+(\check{\kk}\cdot\check{\qq})^2 \check{\Delta}^2/(\check{\epsilon}_f^{\rm eff}(\check{q})/2)^{3/2} +O(k^4)$.
Then $\partial_{\check{\omega}} \check{I}_{11}={\check{\omega}}^2 J +O(1)$, $\partial_{\check{\omega}} \check{I}_{12}=(\check{q}^2-4) J/2+O(1)$,
$\partial_{\check{\omega}} \check{I}_{22}=[{\check{\omega}}^2- (2\check{\Delta})^2] J+O(1)$, where the
$O(1)$ remain bounded when $(\check{\omega},\check{q})\to(\check{\epsilon}_f^{\rm eff}(\check{q}_{\rm sup}),\check{q}_{\rm sup})$
whereas $J=\int \frac{d^3\check{k}}{4\pi} [(\check{\epsilon}_{f,\kk+\qq/2}+\check{\epsilon}_{f,\kk-\qq/2}-\check{\omega})
\,\check{\epsilon}_f^{\rm eff}(\check{q})]^{-2}$ diverges. Similarly, $\partial_{\check{q}}\check{I}_{11}=
[-\frac{\mathrm{d}}{\mathrm{d}\check{q}}\check{\epsilon}_f^{\rm eff}(\check{q})] {\check{\omega}}^2 J +O(1)$,
$\partial_{\check{q}}\check{I}_{12}=[-\frac{\mathrm{d}}{\mathrm{d}\check{q}}\check{\epsilon}_f^{\rm eff}(\check{q})]
(\check{q}^2-4) J/2+O(1)$ and $\partial_{\check{q}}\check{I}_{22}=
[-\frac{\mathrm{d}}{\mathrm{d}\check{q}}\check{\epsilon}_f^{\rm eff}(\check{q})] [{\check{\omega}}^2- (2\check{\Delta})^2] J+O(1)$.
Since $\frac{\check{\omega}^2}{\check{I}_{11}}+\frac{{\check{\omega}}^2- (2\check{\Delta})^2}{\check{I}_{22}}-
\frac{\check{q}^2-4}{\check{I}_{12}}$ does not tend to zero, one obtains the property (\ref{eq:tangence}).}~:
\be
\frac{\mathrm{d}}{\mathrm{d}q} \epsilon_b(q_{\rm sup}) =\frac{\mathrm{d}}{\mathrm{d}q} \epsilon_f^{\rm eff}(q_{\rm sup}),
\ \mbox {so that}\  F_{\!b}(\check{q}_{\rm sup}) = F_{\!f}(\check{q}_{\rm sup}).
\label{eq:tangence}
\ee
On the contrary, for $q<q_{\rm sup}$, the (negative) energy deviation $\epsilon_b(q)-\epsilon_f^{\mathrm{eff}}(q)$ has a positive derivative
so that $F_{\!b}(\check{q})<F_{\!f}(\check{q})$. 
Then, close to the $B_1-F$ boundary, the functions $F_{\!b}(\check{q})$ et $F_{\!f}(\check{q})$ decrease and converge to 
a {\sl common} limit $F_{\!f}(\check{q}_{\rm sup})$ when $q\to q_{\rm sup}^-$, 
the former being less than the latter that goes one decreasing beyond $\check{q}_{\rm sup}$, see figure~\ref{fig:ferm}c.
One can then show graphically that the $F-B_1$ boundary is reached at $\alpha=F_{\!f}(\check{q}_{\rm sup})$, with $q_{0,b}=q_{0,f}=q_{\rm sup}$,
and with $v_b(q_{0,b})=v_f(q_{0,f})$ as it should be: this is a second order transition for $v_c^{L}$ according to (\ref{eq:diffv_alpha});
as a consequence, the $F-B_1$ and $B_{q_0}-B_{q_{\rm sup}}$ boundaries exactly coincide and so do the zones in (\ref{eq:egalite_zones})\footnote{
For $\alpha < F_{\!b}(\check{q}_{\rm sup})=F_{\!f}(\check{q}_{\rm sup})$, the minimum of $v_b(q)$, reached at $q_{\rm sup}$, is larger than the one of $v_f(q)$,
reached at $q_{0,f}>q_{\rm sup}$, since $v_f(q_{0,f})< v_f(q_{\rm sup})=v_b(q_{\rm sup})$. For $\alpha>F_{\!f}(\check{q}_{\rm sup})=F_{\!b}(\check{q}_{\rm sup})$,
the minimum of $v_f(q)$, reached at $q_{0,f}< q_{\rm sup}$, is larger than the one of $v_b(q)$, reached at $q_{0,b}< q_{\rm sup}$,
since $v_b(q_{0,b})<v_b(q_{0,f})<v_f(q_{0,f})$.}.

The remaining part is more straightforward. At the $B_1-B_2$ boundary, $q_{0,b}$ jumps from to positive value $q_0^{\rm min}$
(to the right of the $F_{\!b}(\check{q})$ maximum, see figure \ref{fig:Fbex}b) to the value zero, so that $v_c^{L}$ exhibits a first order transition.
There is a similar scenario at the $F-B_2$ boundary, where $v_f^{\rm (i)}(q_{0,f})=c\equiv v_b(q_{0,b}=0)$, and the absolute minimum location
$q_0^{\rm abs}$ jumps from the value $q_{0,f}\geq 2 k_{\rm min}$ to the value zero.

\section{Critical relative velocity of Bose and Fermi superfluids}
\label{sec:cbe}

It is likely that experimental verification of the Landau critical velocity predicted here will be for
many impurities, rather than one. Since it is desirable to send into the Fermi superfluid 
a homokinetic ensemble of impurities, one is naturally led to use a Bose-Einstein condensate of such impurities.
Interactions among impurities will then be in general significant, as is the case in the reference \cite{Salomon_melange}.
Landau's reasoning must therefore be generalised to the case of a Bose superfluid moving at velocity
$\vv$ inside the Fermi superfluid.

The Bose superfluid is initially at zero temperature in its center-of-mass frame. The arbitrarily weak density-density
interaction between the bosons and the fermions, see section \ref{sec:intro}, creates at least one elementary excitation 
in the Bose superfluid, of momentum $\hbar\qq$ and energy $\epsilon_{B,\qq}{ +}\hbar \qq\cdot\vv$, $\qq\mapsto \epsilon_{B,\qq}=\epsilon_B(q)$
being the dispersion relation for a superfluid at rest\footnote{
This results from the following properties of the unitary transform $T_t(\vv)$ setting the gas into motion as a whole
at velocity $\vv$, that is a Galilean boost at velocity $-\vv$, $T_t(\vv)=\exp[-\mathrm{i}\sum_j t \vv \cdot \pp_j{ /}\hbar]\exp[\mathrm{i}\sum_j m_B \vv
\cdot\rr_j/\hbar]{ \exp[\mathrm{i}tN_B m_B v^2/2]}$, where the sum is over the $N_B$ bosons, of mass $m_B=M$,
position operators $\rr_j$ and momentum operators $\pp_j$:
$T_t^\dagger(\vv) H_B T_t(\vv) =H_B+\vv\cdot \PP_B + N_B m_B v^2/2$  and $T_t^\dagger(\vv) \PP_B T_t(\vv)= \PP_{ B}+
N_B m_B \vv$, where $H_B$ is the Hamiltonian of the bosons and $\PP_B$ is their total momentum operator.
It then remains to compare the energies and momenta of $T_t(\vv) |\Psi_0\rangle$ and $T_t(\vv) |\Psi_1^\kk\rangle$,
where the state vectors $|\Psi_0\rangle$ and $|\Psi_1^\kk\rangle$ represent the superfluid at rest in its ground state
or in presence of an elementary excitation of wavevector $\kk$.}.
Concomitantly, a pair of fermionic excitations of wavevectors $\kk_1$ and $\kk_2$ and of energy $\epsilon_{f,\kk_1}+\epsilon_{f,\kk_2}$,
with $\qq=-(\kk_1+\kk_2)$, or a bosonic excitation of wavevector $-\qq$ and energy $\epsilon_{b,-\qq}$
appears in the Fermi superfluid. This minimal excitation process cannot conserve energy if the relative velocity $v$
of the two superfluids is below the critical Landau velocities
\bea
v_{c,f}^{L}=\inf_\qq v_f(q) \ \ \mbox{with}\ \ v_f(q)=\frac{\epsilon_B(q)
+\epsilon_f^{\rm eff}(q)}{\hbar q} \\
v_{c,b}^{L}=\inf_\qq v_b(q) \ \ \mbox{with}\ \ v_b(q)=\frac{\epsilon_B(q)
+\epsilon_b(q)}{\hbar q},
\eea
that is below $v_c^{L}$, which is the smallest of the two velocities.
In what follows, we shall use the Bogoliubov form
\be
\epsilon_B(q)=\left[\frac{\hbar^2q^2}{2M} \left(\frac{\hbar^2q^2}{2M}+2\mu_B\right)\right]^{1/2}
\ee
where $\mu_B$ is the (positive) chemical potential of the Bose superfluid at rest and $M$ is the mass of a boson.
The previous expressions (\ref{eq:vcfut}) and (\ref{eq:vcb}) correspond, as it should be, to the limiting case $\mu_B\to 0$.

The analysis of the critical velocity $v_{c,f}^{L}$ on the fermionic branch can be done analytically, introducing dimensionless variables
as in equation (\ref{eq:adimf}) and the width $\check{Q}_B$ of  $\check{\epsilon}_B(\check{q})/\check{q}$, given by
\be
\check{Q}_B^2=\frac{2 M}{m} \frac{\mu_B}{|\mu|}=\frac{2\check{\mu}_B}{\alpha}
\ee
One finds identifies the local minima, of zero derivative:
\be
\check{v}_f(\check{q})=\alpha(\check{q}^2+\check{Q}_B^2)^{1/2}+\frac{\check{\epsilon}_f^{\rm eff}(\check{q})}{\check{q}}, \ \ \mbox{so that }\ \ \frac{\mathrm{d}}{\mathrm{d}\check{q}}
\check{v}_f(\check{q}) = \frac{\alpha \check{q}}{(\check{q}^2+\check{Q}_B^2)^{1/2}} -F_{\!f}(\check{q})
\ee
The root of that expression lies on the stationarity branch (iv) [rather than on branch (i)] if and only if the increasing function
$\check{q}\mapsto \alpha\check{q} /(\check{q}^2+\check{Q}_B^2)^{1/2}$ reaches the value $\check{\Delta}/2$  for $\check{q}\in [0,2]$,
that is if and only if
\be
\frac{2\alpha}{(4+\check{Q}_B^2)^{1/2}} \geq \frac{\check{\Delta}}{2},
\label{eq:quand_iv_gagne}
\ee
as can be shown graphically with the help of figure \ref{fig:ferm}{ c}.
The plane $(\check{\Delta},\alpha)$ is thus again split in two domains $F_{\rm (i)}$ and $F_{\rm (iv)}$, and
$\check{v}_{c,f}$ exhibits a second order transition at their boundary.

The analysis of the critical velocity $v_{c,b}^{L}$ on the bosonic branch is performed numerically.
Similarly to the case with one impurity, one finds that the plane $(\check{\Delta},\alpha)$ is split in three domains $B_0$, $B_{q_{\rm sup}}$
and  $B_{q_0}$, depending on where the absolute minimum of $v_{b}(q)$ is located, at the lower endpoint, at the upper endpoint or in
the interior of the existence interval $[0,q_{\mathrm{sup}}]$ of the bosonic branch.
The boundaries only weakly differ from the one for the single impurity, since the Bose chemical potential was taken to be small as compared
to the Fermi energy of the fermions. Note that $v_{c,b}^{L}=c+c_B$ in the whole zone $B_0$, $c$ and $c_B$ being the sound velocities
in the Fermi and Bose superfluids at rest.

The diagram in the plane $(\check{\Delta},\alpha)$ for the global critical velocity $v_{c}^{L}$ is shown in figure \ref{fig:diagvc}b and is described in detail in the caption.
The results and their discussion are close to the single impurity case, see the previous section.
We just report a noticeable difference: the dotted line obeying the equation $v_{c,f}^{L}=c+c_B$  
no longer gives a good approximation to the $B_1-B_2$ boundary, except close to the triple point.

\section{Conclusion}
\label{sec:conclusion}

We have extended Landau's calculation of the critical velocity in a non-polarized Fermi superfluid 
to the case where the moving object is (a) an impurity of finite  mass $M$, and (b) a superfluid of such bosonic impurities, 
taking into account the BCS pair breaking excitations of the Fermi superfluid (fermionic excitation branch) 
and the RPA excitation of the pair center-of-mass motion (bosonic excitation branch) as in reference \cite{CKS}.

When the chemical potential of the fermions is negative, $\mu<0$, we find that the critical velocity is determined
by the phononic part of the bosonic excitation branch and is therefore simply (a) the sound velocity $c$ in the Fermi superfluid,
or (b) the sum $c+c_B$, where $c_B$ is the sound velocity in the Bose superfluid at rest.

When the chemical potential of the fermions is positive, $\mu>0$, these results only apply to some zone $B_2$
in the plane $(\Delta/\mu, m/M)$, where $\Delta$ is the gap and $m$ the mass of a particle in the Fermi superfluid.
For (a)  as well as for (b), at least when the chemical potential $\mu_B$ of the bosons at rest 
is small as compared to the Fermi energy of the fermions,
there exist two other zones, a zone $B_1$ where the critical velocity is determined by the intermediate (non phononic) part of the
bosonic excitation branch,  and a zone $F$ where the critical velocity is the one $v^{L}_{c,f}$ on the fermionic excitation branch.
The critical velocity has a discontinuous second order differential at the $F-B_1$ boundary, and a discontinuous first order differential at the
$F-B_2$ and $B_1-B_2$ boundaries. The three boundaries merge at a triple point. The $F-B_2$ boundary exactly obeys the equation $v^{L}_{c,f}=c$ 
[case (a)] or $v^{L}_{c,f}=c+c_B$ [case (b)]. 
Similarly, on the $B_1-B_2$ boundary, the critical velocities coming from the phononic
part and the intermediate part of the bosonic excitation branch are exactly equal; in the case (a), one gets a good approximation
by solving the simpler equation $v^{L}_{c,f}=c$, because the critical velocity in $B_1$ is actually very close to $v^{L}_{c,f}$;
in the case (b), this is not an as good approximation, except in the vicinity of the triple point.
Last, the $F-B_1$ boundary is exactly on the line $v_{c,f}^L=v_f(q_{\rm sup})$, where the function $v_f(q)$ is minimal at
the maximal wavenumber $q_{\rm sup}$ of the bosonic excitation excitation branch,  and it can be well approximated both for (a) and (b) 
by a portion of the line of discontinuity of the second order differential of $v_{c,f}^{L}$; this line is
given by the equation $m/M=\Delta/(2\mu)$ for the case (a), and by assuming equality rather than inequality in
 (\ref{eq:quand_iv_gagne}), for the case (b).

These predictions may be verified experimentally with the mixture of superfluids of bosonic ${}^7$Li and fermionic ${}^6$Li
isotopes of lithium that was recently prepared at ENS \cite{Salomon_melange}.
For example, the predicted first order transition at the $B_1-B_2$ boundary may be revealed through a variation of the scattering length
of the opposite spin fermions (this will change $\check{\Delta}$) with a Feshbach resonance; it remains to measure the corresponding
critical velocity and to check that it has a kink as a function of the interaction strength, at the crossing point of
the $B_1-B_2$ boundary. The fixed value $m/M\simeq 6/7$ of the mass ratio does not allow, however, to cross the other boundaries.

It may be possible to extend our theoretical study to what was directly measured in reference \cite{Salomon_melange},
that is the damping rate of the Bose superfluid oscillations within the harmonically trapped Fermi superfluid, including
possible non-zero temperature effects. It also remains to see if the boson-fermion interaction is indeed so weak
that one can make an analysis \`a la Landau, restricting to the minimal number of elementary excitations, and 
obtain the same energy barrier (preventing damping of the motion of the impurities in the Fermi superfluid)
as in the experiment. We hope that these questions will provide some inspiration for future works, theoretical or experimental.

\section*{Acknowledgements}
Our group is also affiliated to IFRAF. We acknowledge financial support from the Nano-K DIM (ATOMIX project)
and from the Institut de France (Louis D.\ prize). We thank  the members of the ``cold fermions" group, as well as Claude Cohen-Tannoudji, Franck Lalo\"e
and Xavier Leyronas for useful discussions.

\end{document}